\documentclass[12pt, prd, showpacs]{revtex4}
\usepackage{amssymb}
\usepackage{amsmath}
\usepackage{graphicx}

\begin{document}

\title{Full set of scenarios of high energy collision in the Schwarzschild
background and kinematic censorship}
\author{A. V. Toporensky}
\affiliation{Centre for Cosmology and Science Popularization (CCSP),}
\affiliation{SGT University, Gurugram, Delhi-NCR, Haryana 122505, India}
\email{atoporensky@gmail.com}
\author{P. Gusin}
\affiliation{Faculty of Basic Problems of Technology (Wroclaw), Wroclaw University of
Science and Technology, 50-370 Wroclaw, Poland}
\author{A. Radosz}
\affiliation{Faculty of Basic Problems of Technology (Wroclaw), Wroclaw University of
Science and Technology, 50-370 Wroclaw, Poland}
\author{O. B. Zaslavskii}
\affiliation{Department of Physics and Technology, Kharkov V.N. Karazin National
University, 4 Svoboda Square, Kharkov 61022, Ukraine}
\email{zaslav@ukr.net }

\begin{abstract}
We consider collisions between particles moving freely in the Schwarzschild
background. We suggest classification of scenarios that lead to unbounded
energy output in the center of mass frame. Although this is forbidden in the
exterior of a black hole, the phenomenon becomes possible with account of
white hole region and mirror universe in the complete space-time diagram.
Also, we scrutinize collisions in the vicinity of the bifurcation point and
reveal the role of resting observers in the inner region beyond the horizon.
In some scenarios $E_{c.m.}$ not only diverges but becomes seemingly
infinite that is, as a matter of fact, impossible according to the principle
of so-called kinematic censorship. We resolve a corresponding paradox and
argue, why this principle remains valid.
\end{abstract}

\maketitle

\section{Introduction}

Since 2009, a lot of attention was focused on high energy particle
collisions near black holes due to findings of \cite{ban} called the Ba\~{n}%
ados-Silk-West (BSW) effect. According to it, collisions of two particles
can lead, under some conditions, to the unbounded growth of the energy
output in their center of mass frame $E_{c.m.}$ in the point of collision.
These conditions considered in \cite{ban} included (i) the extremal horizon,
(ii) fine-tuning of parameters of one of the particles. In the case of
rotating black holes, fine-tuning relates the energy and angular momentum.
Although in the pioneering work \cite{ban} only the Kerr metric was
considered, actually the BSW effect is inherent to generic rotating black
holes and for a wide class of horizons \cite{prd}, \cite{gen}.

As in the Schwarzschild metric there is no rotation, the original BSW effect
is impossible, so $E_{c.m.}$ remains limited. Collision of two particles
with mass $m$ near the horizon can give $E_{c.m.}=m\sqrt{5}$ only \cite{baus}%
. Meanwhile, such a conclusion implies that the collision occurs outside the
horizon in the outer region. However, the full space-time diagram of the
Schwarzschild black hole covers four qualitatively different regions. This,
apart from "Our Universe", includes the Black Hole and White Hole regions
and also the Mirror Universe region. And, if one takes into account
collisions there, the situation changes. Some results for particular
scenarios of particle collisions in these regions were already obtained in 
\cite{white-black}. The purpose of the present work is to give\textit{\ full 
}classification of high energy particle collisions in the Schwarzschild
background.

Also, we would like to stress that the result \cite{baus} implies (i)
collision in our world outside a black hole, (ii) particles moving along the
geodesic paths. If we relax condition (ii) and choose of one particles to be
in a rest, $E_{c.m.}$ can become as large as one like. However, a resting
particle outside the horizon does certainly not follow a geodesic.
Meanwhile, such particles do exist inside the horizon (we call them $RO$ -
resting observers) and this makes these observers stand out.

In the real world, the mirror universe is absent and cannot appear in the
course of gravitational collapse. There are also arguments that prevent
existence of white holes as physical objects on their own. Hence, why, in
spite of this, the problem under consideration deserves attention?
Motivation comes from several directions. (i) General viewpoint: if a theory
predicts some nontrivial effects, we should scrutinize its consequences as
fully as possible. (ii) Does such a type of collisions contribute to the
instability of white holes and thus explain (at least partially) why they do
not exist in nature? This reverses the logic: instead of rejecting the
subject of investigation because of instability of a corresponding object,
we try to understand why such an instability can happen and what is the role
of the phenomenon under discussion. (iii) Building a general scheme can be
viewed as a first step towards constructing a similar one for black holes
possessing an inner \ horizon thus helping to understand the similarity (or
difference) between instability of white holes and that of inner horizons.
Thus we essentially generalize observations about high energy particle
collisions due to white holes noticed in \cite{gpwhite}.

The paper is organized as follows. To make presentation self-contained, in
Sec. \ref{str} we list the metric and we give the main types of geodesics
relevant for our problem. In Sec. \ref{sce} we describe scenarios of
collision with high energy outcome in which none of particles passes through
the bifurcation point. Further, we analyze different group of scenarios and
discuss the particular scenarios of collisions. In Sec. \ref{lem} we discuss
some properties of scenarios under consideration using the Lema\^{\i}tre
frame. In Sec. \ref{mar} we consider the case when one of the colliding
particles, whose trajectory originates in a white hole region,\textit{\ }is
a so-called resting observer ($RO$), so it passes through the bifurcation
point near which it collides with another massive particle. In Sec. \ref%
{enear} we consider collisions near the bifurcation point in which $RO$ do
not participate.\ In Sec. \ref{il} we discuss seeming violation of the
principle of kinematic censorship and explain the solution of the
corresponding paradox. In Sec. \ref{massless} we derive general formulas for
the case when a resting observer collides with a massless particle. In Sec. %
\ref{kruskal} we discuss how the picture looks like in the Kruskal-Szekeres
coordinates. In Sec. \ref{nearhor} we consider near-horizon collisions of
such particles, discuss different limits and show that the principle of
so-called kinematic censorship (in any event energy cannot be infinite) is
fulfilled. In Sec. \ref{concl} we summarize the results.

\section{Causal structure and notations\label{str}}

In the standard coordinates the Schwarzschild metric reads%
\begin{equation}
ds^{2}=-fdt^{2}+\frac{dr^{2}}{f}+r^{2}(d\theta ^{2}+\sin ^{2}\theta d\phi
^{2})\text{,}  \label{1met}
\end{equation}%
where $f=1-\frac{2M}{r}$, $M$ is the black hole mass. Such a coordinate
systems spoils at the horizon $r=2M$. In the Kruskal-Szekeres coordinates
system that is regular everywhere and fully covers the Schwarzschild
space-time manifold, one distinguishes four subsets referred as $I$ - Our
Universe (OU); $II$ - Black Hole interior (BH); $III$ - Mirror Universe
(MU); $IV$ - White Hole (WH). T(see Fig.1). They are separated by the
horizons: F (Future)-Horizon (FH), composed of the left branch of the WH's
horizon and future BH's horizon and P(Past)-Horizon (PH), composed of the
right branch of the WH's one and the BH's horizon (which plays a role of
future horizon in MU). Alternatively, one can use classification to R and T
regions \cite{novrt}: $I=R+$, $II=T-$, $III=R-$, $IV=T+$.

\begin{figure}
    \centering
    \includegraphics[width=1\linewidth]{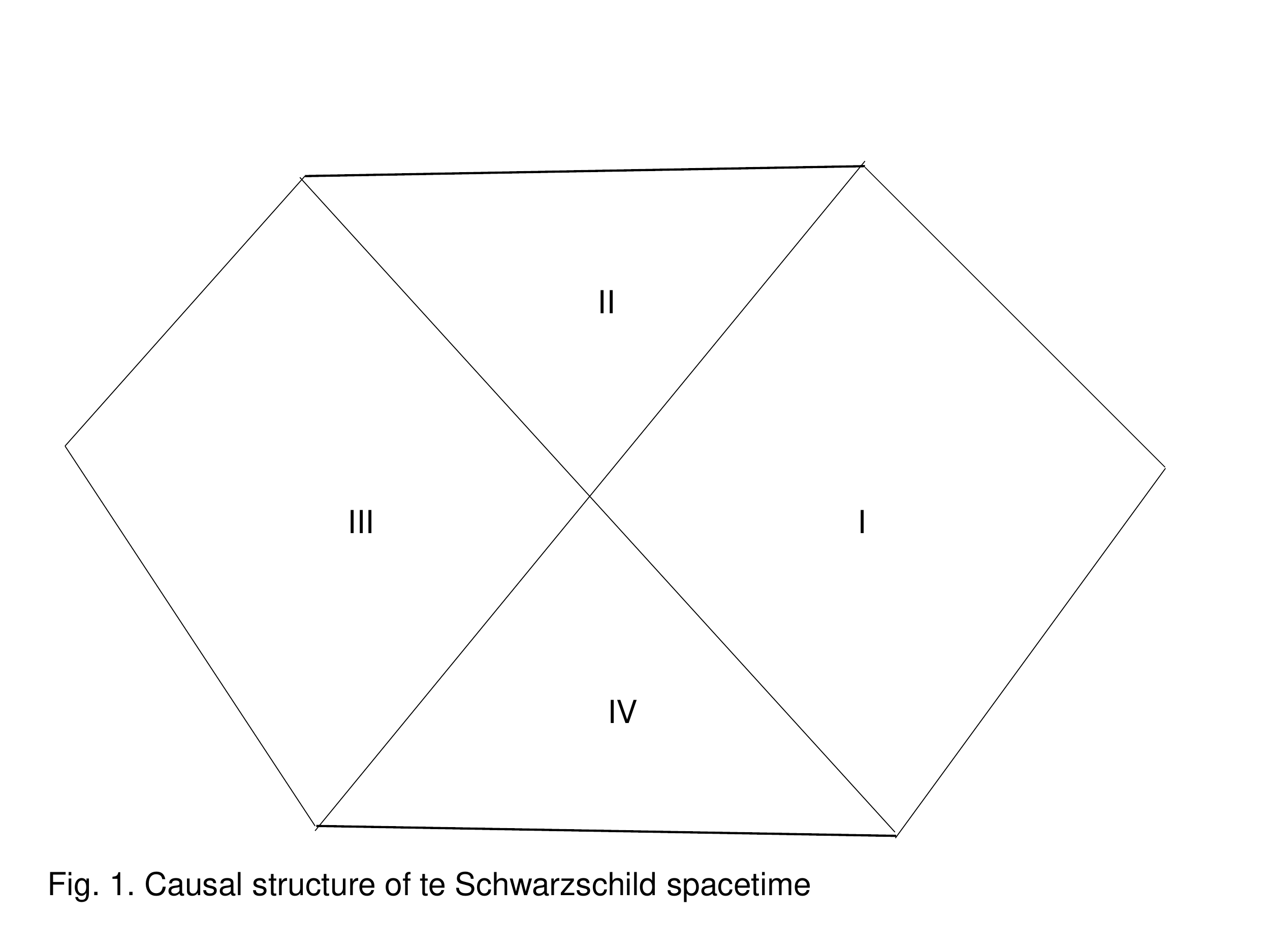}
    \caption{Causal structure of the Schwarzschild space-time.}
\end{figure}

\subsection{Time-like geodesics - general remarks}

Let us begin with consideration of the candidates for the collisions of the
highest energy outcome, i.e. time-like geodesics confined to the $t-r$
hyperplane within the Schwarzschild spacetime (region $I$ - see Fig.1). It
is well-known that, despite the fact that the Schwarzschild coordinates
reveal singular behavior on the black hole's horizon, they might be applied
in the interior of a black hole, i.e. in the region $II$ \cite{nov61}. In
region $I$ $\left( OU\text{ - our Universe}\right) $, variable $t$ is a
temporal coordinate, $dt>0$ and $r$ is a spatial coordinate. However, in
region $II,$ variables $t,$ and $r$ interchange their roles: $t\in \left(
-\infty ,\infty \right) $ becomes space-like coordinate and $r\in \left(
2M,0\right) $ becomes time-like coordinate, $dr<0$ (see also below). Taking
this into account one can find the velocity vector $u$ of the temporal
geodesics belonging to the $t-r$ hyperplane from the geodesic equations as
follows:

\begin{equation}
u=\sigma \frac{\varepsilon }{f}\partial _{t}+\sigma ^{\prime }\sqrt{%
\varepsilon ^{2}-f}\partial _{r},  \label{1}
\end{equation}%
where, $f=1-\frac{2M}{r}$, $\sigma ,\sigma ^{\prime }=\pm 1$ (and we take $%
\varepsilon \geq 0$ by definition). The quantity $\varepsilon $ has the
meaning of the absolute value of energy (per unit mass) in the R-region or
momentum (per unit mas) in the T-region.

In terms of coordinates, $r=-T$ in the black hole region $II$ and $r=+T$ in
the white hole region $IV$. The geodesics can be classified as follows.

\begin{enumerate}
\item $I$ $\left( f>0,dt>0\right) .$ In $OU$, the forward-in-time condition, 
$dt>0$ imposes $\sigma =+1$ and $\sigma ^{\prime }=\pm 1$ that corresponds
to the "out" $\left( +\right) $ and "in" $\left( -\right) ,$ geodesics.

\item $II$ $\left( f<0,dr<0\right) .$ In the black hole interior $BH$, the
forward-in-time condition, $dr<0,$ implies that $\sigma ^{\prime }=-1,$ and $%
\sigma =\pm 1$ represents temporal geodesics with the motion along $t-$axis
(which is spatial) in both directions. Geodesic that enters region $II$ from 
$I~$has $\sigma =+1$ whereas the one that enters $II$ from $III$ has $\sigma
=-1$.\ \ \ \ \ \ \ \ \ \ \ \ \ \ \ \ \ \ \ \ \ \ \ \ \ \ \ \ \ \ \ \ \ \ \ \
\ \ \ \ \ \ \ \ \ \ \ \ \ \ \ \ \ \ \ \ 

\item $III\left( f>0,dt<0\right) .$ In the mirror universe $MU,$ the
forward-in-time condition imposes $\sigma =-1$ (coordinate time $t$ is
changing from $+\infty $ to $-\infty $) and $\sigma ^{\prime }=\pm 1.$

\item $IV\left( f<0,dr>0\right) .$ In the interior of the $WH$ , $\sigma
^{\prime }=+1$ and $\sigma =\pm 1$ represents geodesics entering $OU$ $%
\left( +\right) $, or geodesics entering $MU$ $\left( -\right) .$
\end{enumerate}

A special class of temporal geodesics are those representing resting
observers ($RO$) inside $BH$ ($II$) and inside $WH$ ($IV$) \cite{dor}. Such
observers have no analogue in region $I$ (or $III$) where a particle at rest
requires a force to keep it fixed.

They are resting on $t-$ axis, so for them $\varepsilon =0$ and their
four-velocity vector is

\begin{equation}
u=\left\{ 
\begin{array}{cc}
\sigma ^{\prime }=+1,\text{ \ \ \ }\sqrt{-f}\partial _{r} & WH \\ 
\sigma ^{\prime }=-1,\text{ }-\sqrt{-f}\partial _{r} & BH%
\end{array}%
\right\} .
\end{equation}

The temporal geodesics confined within the $t-r$ hyperplane may be easily
expressed in the other coordinate systems, in particular in the
Kruskal-Szekeres (KS) one (that covers the full set $I-IV$), via
transformation of the velocity vector (\ref{1}) from the Schwarzschild's
into the KS's system of coordinates. One verifies then the smooth behavior
of the velocity vector of the particle while crossing the horizon (see
below).

See also more detailed classification that takes into account dynamic
properties including the sign of energy-momentum in \cite{genrad}.

\subsection{Light-like geodesics}

One can apply the above analysis to the categorization of the light-like
geodesics confined to the $t-r$ hyperplane. They have an especially
appealing feature in the KS coordinates: they constitute the system of
mutually perpendicular straight lines, parallel and perpendicular to the F
and P horizons (see Fig.2).

\begin{figure}
    \centering
    \includegraphics[width=1\linewidth]{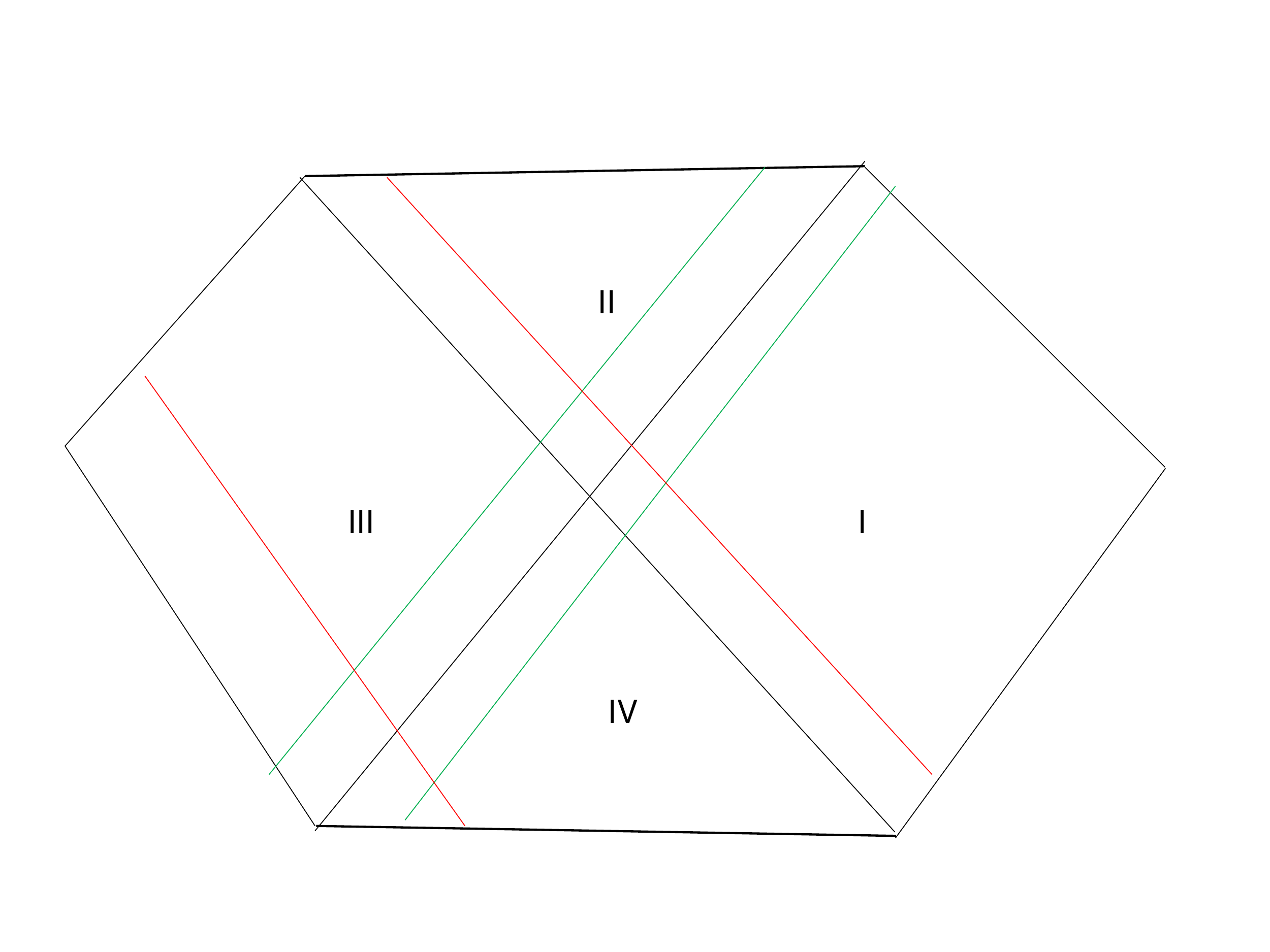}
    \caption{Light-like geodesics in the Schwarzschild space-time.}
\end{figure}

In the analogy to the velocity vector (\ref{1}) of the time-like geodesics,
one can represent the wave vector of light-like geodesics:

\begin{equation}
k=\sigma \frac{\omega _{0}}{f}\partial _{t}+\sigma ^{\prime }\omega
_{0}\partial _{r},  \label{k}
\end{equation}%
where one takes $\omega _{0}>0$.

In the regions $I-IV$ the wave vector (\ref{k}) is associated with the same
set of $(\sigma ,\sigma ^{\prime })$ as the velocity vector of the time-like
geodesics described above.

A separate question arises when a light ray propagates along the horizon -
such a class of null geodesics cannot be described in terms of
Schwarzschild-like coordinates $t-r$ as those are applied outside or inside
of the horizon,\ but not \textit{on} the horizon. Applying the KS
coordinates, one finds that the parameter $\omega _{0}$ has to vanish, $%
\omega _{0}=0$ \cite{along}.

\section{Scenarios of collisions\label{sce}}

If the two particles collide, the energy in the center of mass frame $%
E_{c.m.}$ is defined according to%
\begin{equation}
E_{c.m}^{2}=-P_{\mu }P^{\mu }=m_{1}^{2}+m_{2}^{2}+2m_{1}m_{2}\gamma \text{.}
\end{equation}%
where $P^{\mu }=m_{1}u_{1}^{\mu }+m_{2}u_{2}^{\mu }$, $\gamma =-u_{1\mu
}u_{2}^{\mu }$ is the Lorentz gamma factor of relative motion. For pure
radial motion (with zero angular momenta), taking into account (\ref{1}),
one finds for geodesic motion

\begin{equation}
\gamma =\frac{\sigma _{1}\sigma _{2}\varepsilon _{1}\varepsilon _{2}-\sigma
_{1}^{\prime }\sigma _{2}^{\prime }\sqrt{\varepsilon _{1}^{2}-f}\sqrt{%
\varepsilon _{2}^{2}-f}}{f}.  \label{ga}
\end{equation}%
Collision can lead to unbounded growth of $\gamma $, hence, $E_{c.m.}$ if it
occurs near the horizon where $f$ $\ $is small. Inside the horizon, i.e. in
regions $II$ and $IV$, $f<0$.

If particle 2 is massless, this formula is modified:%
\begin{equation}
\gamma =\frac{\varepsilon _{2}}{f}(\sigma _{1}\sigma _{2}\varepsilon
_{1}-\sigma _{1}^{\prime }\sigma _{2}^{\prime }\sqrt{\varepsilon _{1}^{2}-f})%
\text{.}  \label{radf}
\end{equation}

\bigskip Inside the black or white hole region, it is convenient to rewrite (%
\ref{ga}) in the form

\begin{equation}
\gamma =\frac{\sigma _{1}^{\prime }\sigma _{2}^{\prime }\sqrt{\varepsilon
_{1}^{2}+\left\vert f\right\vert }\sqrt{\varepsilon _{2}^{2}+\left\vert
f\right\vert }-\sigma _{1}\sigma _{2}\varepsilon _{1}\varepsilon _{2}}{%
\left\vert f\right\vert }.  \label{gag}
\end{equation}

If particle 2 is massless,%
\begin{equation}
\gamma =\varepsilon _{2}\frac{(\sigma _{1}^{\prime }\sigma _{2}^{\prime }%
\sqrt{\varepsilon _{1}^{2}+\left\vert f\right\vert }-\sigma _{1}\sigma
_{2}\varepsilon _{1})}{\left\vert f\right\vert }.  \label{radg}
\end{equation}

Below, we mainly concentrate on collisions of massive particles, unless
opposite is stated explicitly.

Near the horizon,%
\begin{equation}
\gamma \approx \frac{2\varepsilon _{1}\varepsilon _{2}}{\left\vert
f\right\vert },  \label{gad}
\end{equation}%
provided signs of sigmas are chosen properly (see below).

Apart from small $\left\vert f\right\vert $, one more condition is required
for high energy collision to occur. The numerator should not vanish on the
horizon, otherwise its small value would compensate small $\left\vert
f\right\vert $. This condition (the so-called head-on collision) is valid in
the following cases:

\begin{enumerate}
\item $OU$ $\left( I\right) $ : $f\rightarrow +0$, $\sigma _{1}=\sigma
_{2}=1;$ $\ \ \ \ \ \ \ \sigma _{1}^{\prime }\sigma _{2}^{\prime }=-1$

\item $BH$ $\left( II\right) $ : $f\rightarrow -0$, $\sigma _{1}^{\prime
}=\sigma _{2}^{\prime }=-1;$ $\ \ \ \ \sigma _{1}\sigma _{2}=-1$

\item $MU$ $\left( III\right) $ : $f\rightarrow +0$, $\sigma _{1}=\sigma
_{2}=-1;$ $\ \ \sigma _{1}^{\prime }\sigma _{2}^{\prime }=-1$

\item $WH$ $\left( IV\right) $ : $f\rightarrow -0$, $\sigma _{1}^{\prime
}=\sigma _{2}^{\prime }=1;$ $\ \ \ \ \ \sigma _{1}\sigma _{2}=-1,$

where, we indicated the region in which$\ $the collision occurred.
Therefore, as there are four distinct regions $I-IV$, each separated by two
horizons, there are eight types of high energy head-on collisions, two of
them (close to each of the two horizons) at each of the four regions. The
corresponding details are collected in Table 1. In columns P1, P2 we
indicate a region where a particle originates from. In doing so, we imply
that collision does not happen in the bifurcation point BP (we will use this
term although, strictly speaking, it is a bifurcation sphere) or its
vicinity.
\end{enumerate}

\begin{tabular}{|l|l|l|l|l|l|l|}
\hline
Scenario & P 1 & P 2 & Point of collision & $\sigma $ & $\sigma ^{\prime }$
& $f$ \\ \hline
1 & WH & R+ & Near right white horizon in R+ & $\sigma _{1}=1,$ $\sigma
_{2}=1$ & $\sigma _{1}^{\prime }=1$, $\sigma _{2}^{\prime }=-1$ & $>0$ \\ 
\hline
2 & WH & R+ & Near right black horizon in R+ & $\sigma _{1}=1$, $\sigma
_{2}=1$ & $\sigma _{1}^{\prime }=1$, $\sigma _{2}^{\prime }=-1$ & $>0$ \\ 
\hline
3 & R- & R+ & Near right black horizon in T- & $\sigma _{1}=-1$, $\sigma
_{2}=1$ & $\sigma _{1}^{\prime }=-1$, $\sigma _{2}^{\prime }=-1$ & $<0$ \\ 
\hline
4 & R- & R+ & Near left black horizon in T- & $\sigma _{1}=-1$, $\sigma
_{2}=1$ & $\sigma _{1}^{\prime }=-1$, $\sigma _{2}^{\prime }=-1$ & $<0$ \\ 
\hline
5 & R- & WH & Near left \ white horizon in R- & $\sigma _{1}=-1,$ $\sigma
_{2}=-1$ & $\sigma _{1}^{\prime }=-1$, $\sigma _{2}^{\prime }=1$ & $>0$ \\ 
\hline
6 & R- & WH & Near left \ black horizon in R- & $\sigma _{1}=-1,$ $\sigma
_{2}=-1$ & $\sigma _{1}^{\prime }=-1$, $\sigma _{2}^{\prime }=1$ & $>0$ \\ 
\hline
7 & WH & WH & Near left white horizon in T+ & $\sigma _{1}=1$, $\sigma
_{2}=-1$ & $\sigma _{1}^{\prime }=1$, $\sigma _{2}^{\prime }=1$ & $<0$ \\ 
\hline
8 & WH & WH & Near right white horizon in T+ & $\sigma _{1}=1$, $\sigma
_{2}=-1$ & $\sigma _{1}^{\prime }=1$, $\sigma _{2}^{\prime }=1$ & $<0$ \\ 
\hline
\end{tabular}

Table 1. Classification of high energy collisions not including the
bifurcation point and its vicinity (Pi stands for "Particle 1/2").

Scenarios 1 and 2 as well as 3 and 4 are equivalent dynamically (they have
the same set of parameters $\sigma ,\sigma ^{\prime }$) but they are not
equivalent kinematically and geometrically (they are realized in different
regions of space-time). In scenario 3 particle 1 passes close to the
bifurcation point (but not through it), in scenario 4 this happens to
particle 2. However, in both cases collision itself occurs far from the
bifurcation point.

Below we put figures describing different scenarios, with the reservation
that they have a schematically character, so for simplicity we depict
conditionally trajectories of massive particles by straight lines.

\begin{figure}
    \centering
    \includegraphics[width=1\linewidth]{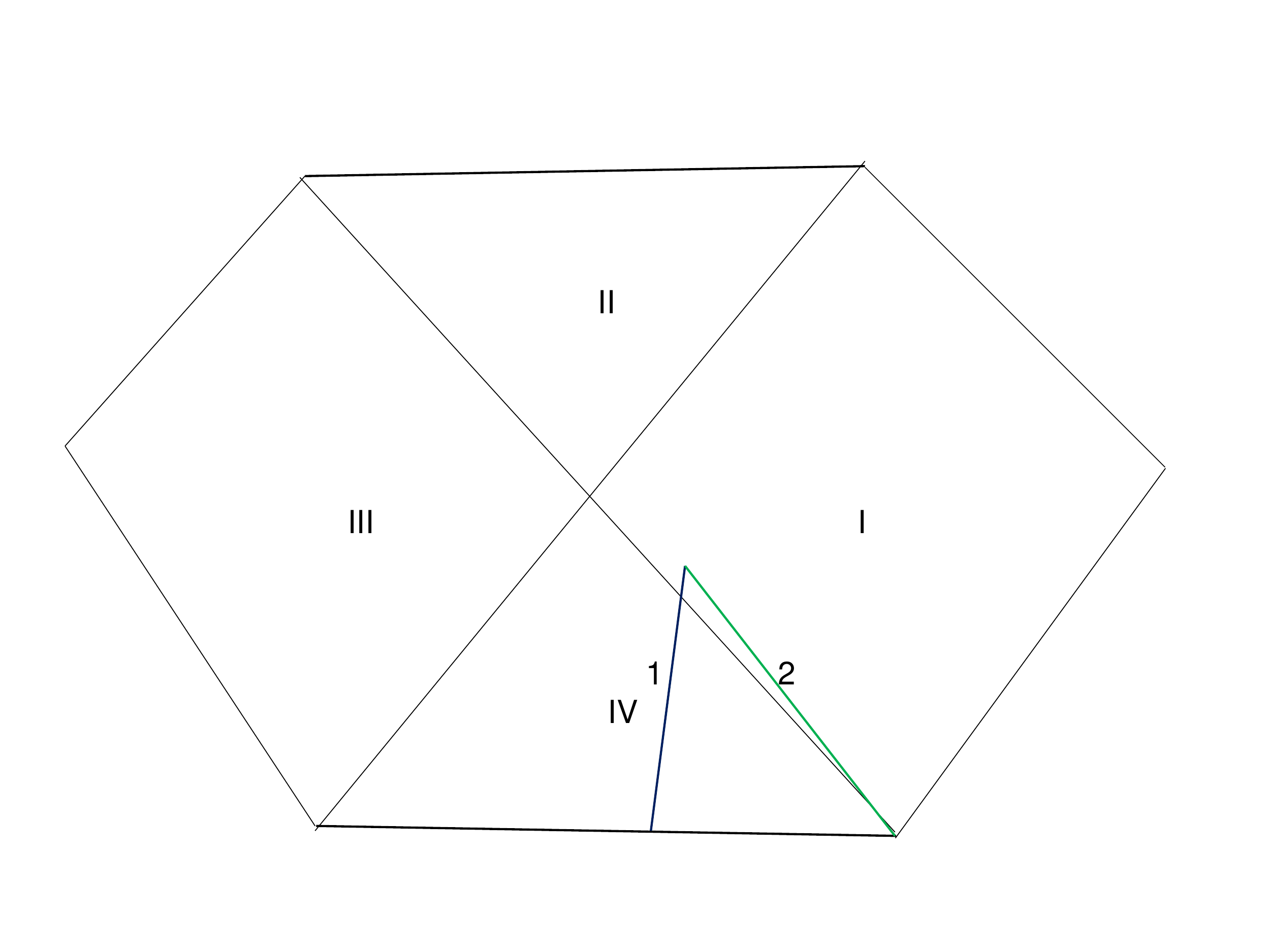}
    \caption{Scenario 1}
\end{figure}

\begin{figure}
    \centering
    \includegraphics[width=1\linewidth]{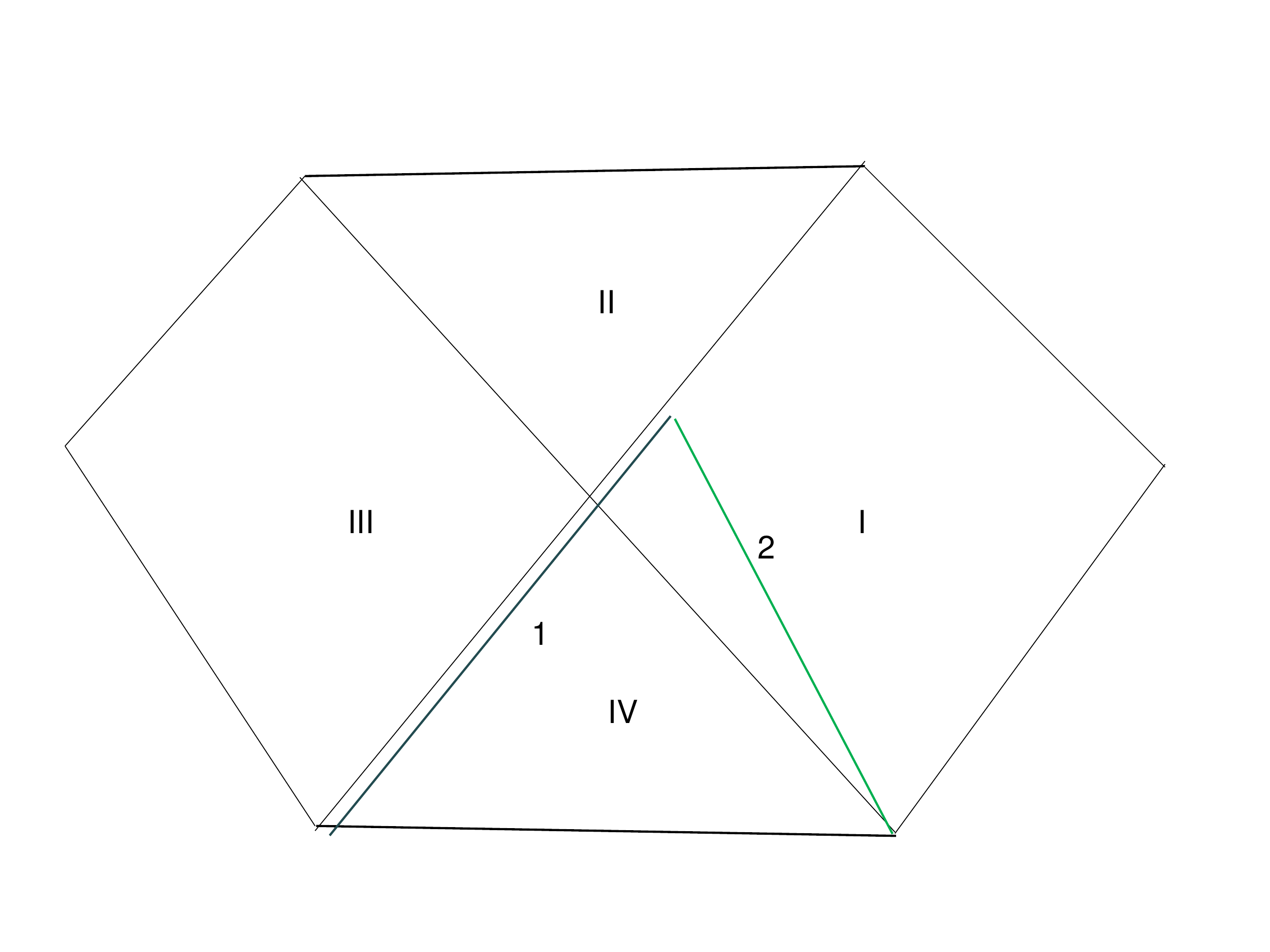}
    \caption{Scenario 2}
\end{figure}

Kinematic conditions of scenarios 1 and 2 were analyzed in \cite{white-black}%
.

\begin{figure}
    \centering
    \includegraphics[width=1\linewidth]{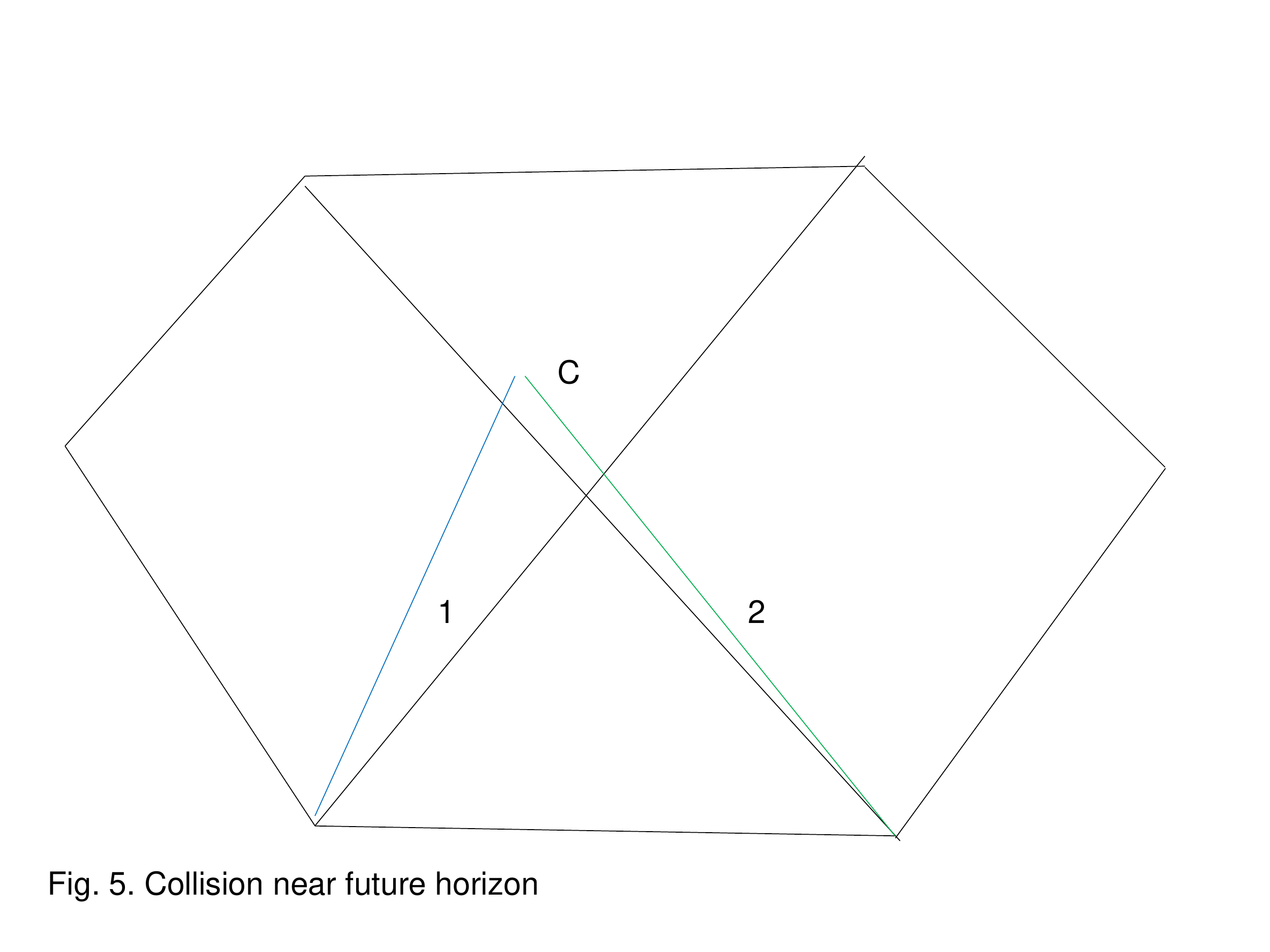}
    \caption{Scenario 3.}
\end{figure}

\begin{figure}
    \centering
    \includegraphics[width=1\linewidth]{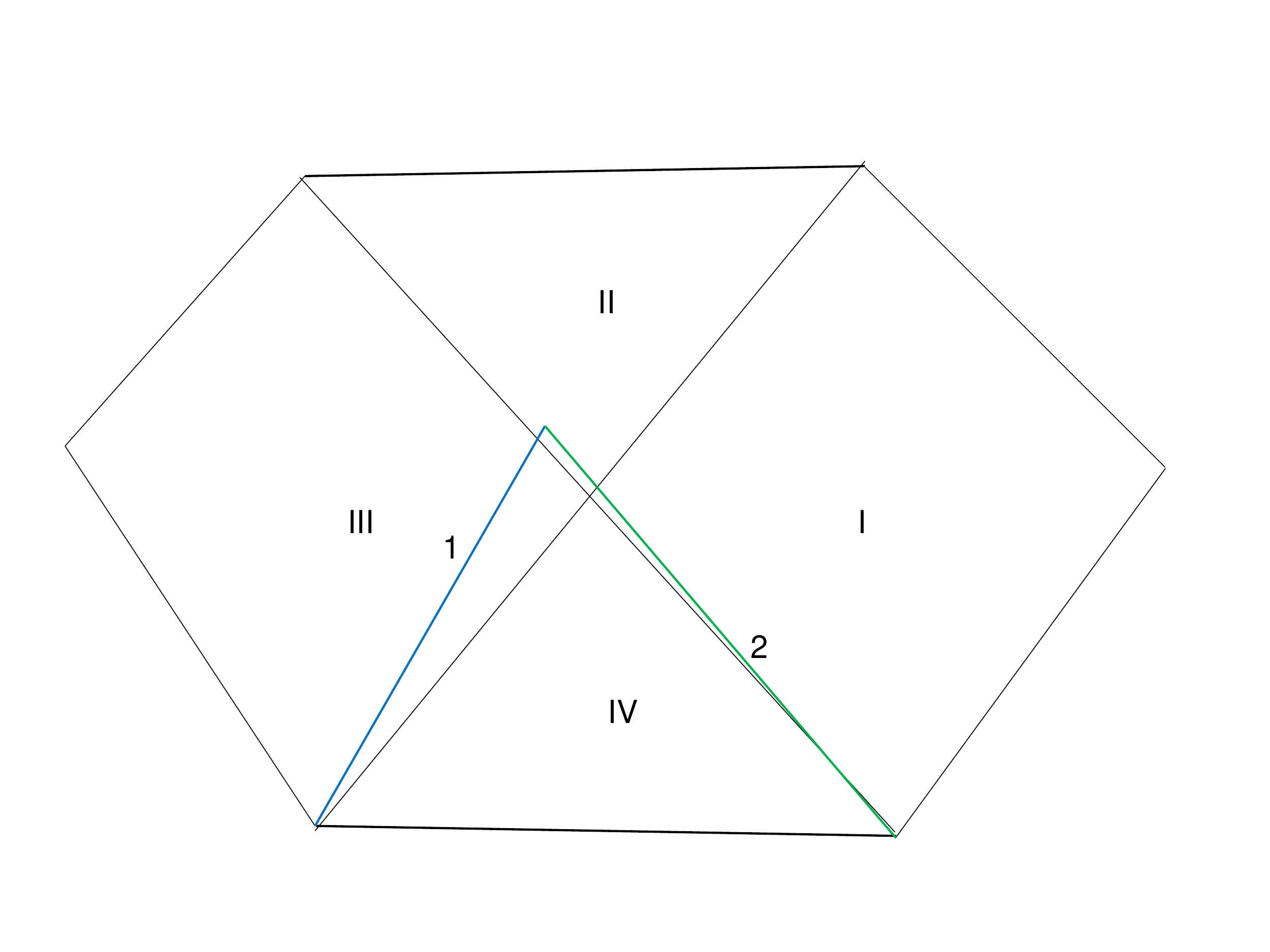}
    \caption{Scenario 4.}
\end{figure}

\begin{figure}
    \centering
    \includegraphics[width=1\linewidth]{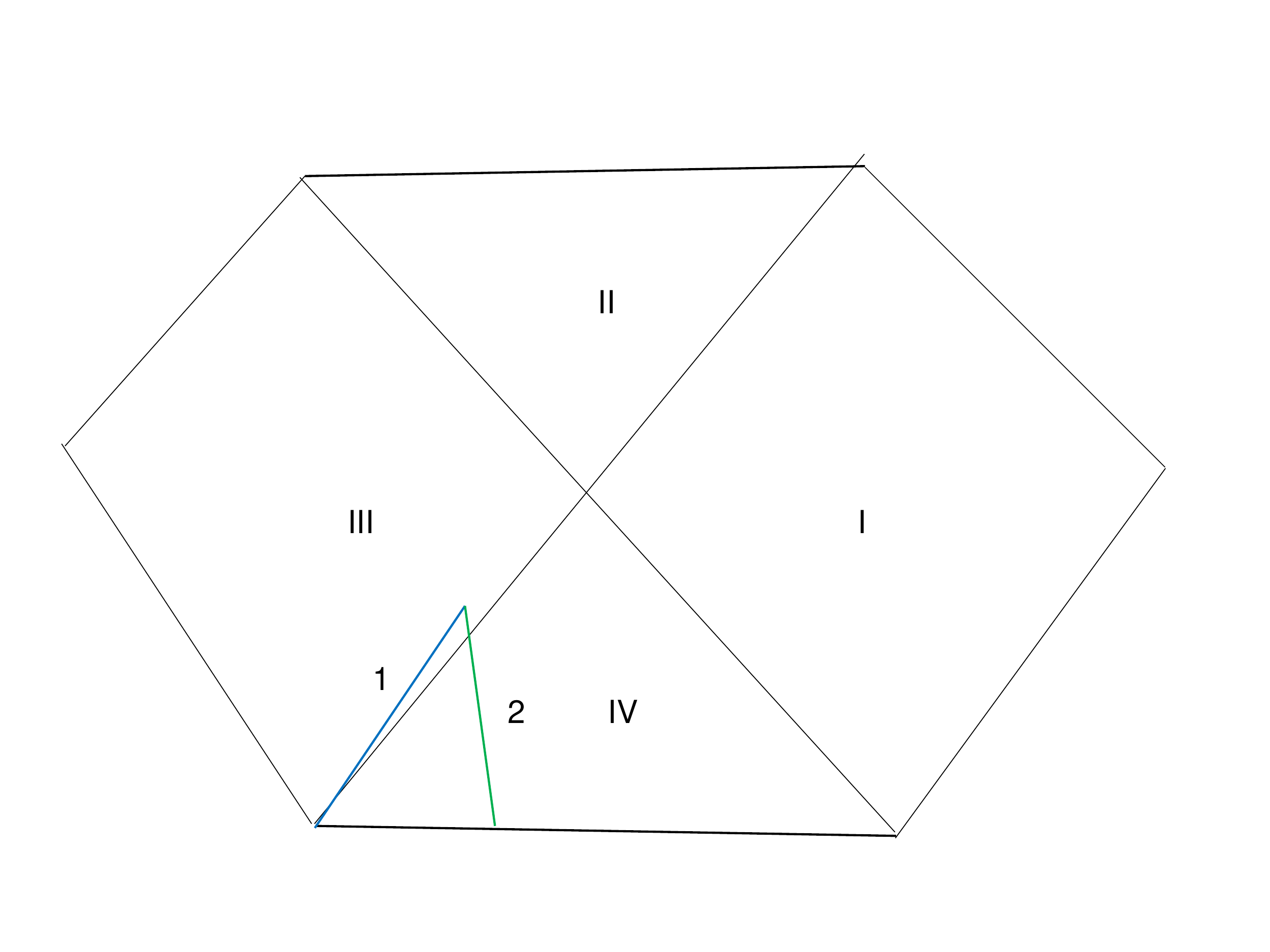}
    \caption{Scenario 5.}
\end{figure}

\begin{figure}
    \centering
    \includegraphics[width=1\linewidth]{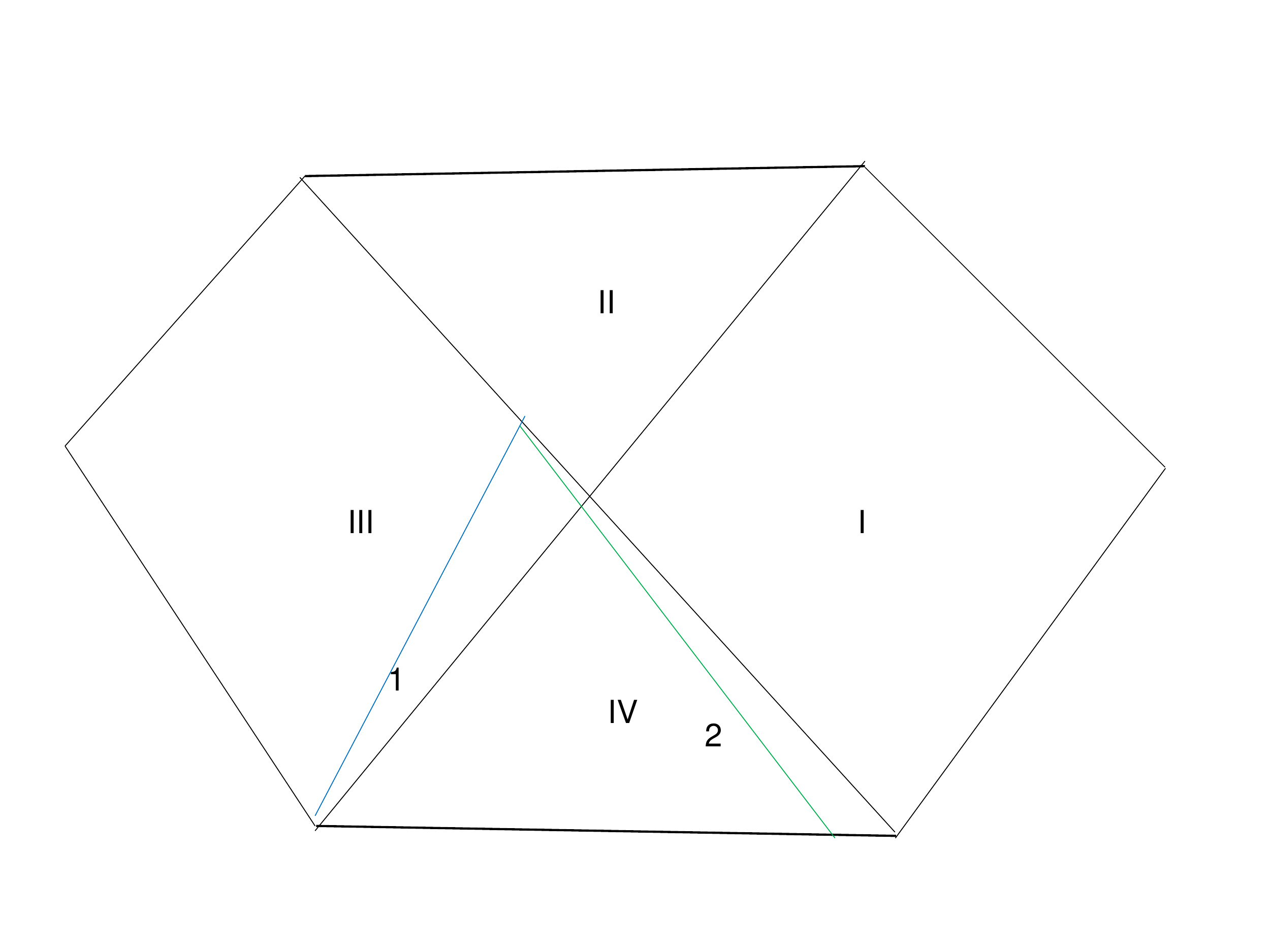}
    \caption{Scenario 6.}
\end{figure}

Scenario 1 and 2 are equivalent, correspondingly, to scenarios 1 and 2 in 
\cite{white-black}.

\subsection{General features of scenarios}

High energy particle collision (with unbounded $\gamma $) requires special
conditions. The corresponding scenario fall in the following classes.

I. Both energies $\varepsilon _{1,}\varepsilon _{2}=O(1)$. II. For one of
particles $\varepsilon =0$ or it is very small. This is direct analogue of
critical (or near-critical) particles typical of collisions in the
background of rotating black holes \cite{ban}, \cite{prd}. In turn, case II
is divided to two subcases. IIa: collision occurs in the vicinity of the
horizon. IIb: collision occurs exactly on the horizon. For case I this is
impossible since this would violate the principle of kinematic censorship 
\cite{cens}. According to it, $\gamma $ cannot be infinite literally in any
event. Below, we will discuss in detail this principle and how it forbids
some scenarios.

For a while, we simply put the formula for $\gamma $ if such a collision
happens (i.e. $f=0$ exactly). This occurs if $\sigma _{1}\sigma _{2}=\sigma
_{1}^{\prime }\sigma _{2}^{\prime }=1$ (so particles have the same
direction). Otherwise, $\gamma $ would be infinite, and this would violate
the kinematic censorship. Then, it follows from (\ref{ga}) that in the $R$
region%
\begin{equation}
\gamma =\frac{1}{2}(\frac{\varepsilon _{1}}{\varepsilon _{2}}+\frac{%
\varepsilon _{2}}{\varepsilon _{1}})\text{.}  \label{gae}
\end{equation}

If $\varepsilon _{1}\ll \varepsilon _{2}$ or $\varepsilon _{2}\ll
\varepsilon _{1}$, $\gamma $ can be made as big as one likes but it remains
finite anyway. Formally, it can be formulated as divergence of $%
\lim_{\varepsilon \rightarrow 0}\lim_{f\rightarrow 0}$.

The common features of scenarios of class I are as follows.

1) High energy processes in the R region require head-on collisions, i.e.
the opposite signs of momenta: $\dot{r}_{1}\dot{r}_{2}<0,$ $\sigma
_{1}^{\prime }\sigma _{2}^{\prime }=-1$

2) High energy collision in the T region also require head-on collisions,
but now formulation of this condition is different: $\sigma _{1}\sigma
_{2}=-1$. This is because energy and momentum interchange their role in the
T region, that is a consequence of the interchange of the $t\leftrightarrow
r $ roles.

3) In all cases of head-on-collisions scenarios, $\gamma \sim \left\vert
f\right\vert ^{-1}$.

4) For scenarios of class IIa $\gamma \sim \left\vert f\right\vert ^{-1/2}$.

\subsection{Scenarios 7 and 8}

In the examples considered above, scenarios in which both particles cross
the horizon approaching it along geodesics originating in the same region,
do not lead to high energy outcome\textit{.} However\textit{, }there are two
scenarios, namely 7 and 8, where both geodesics originate in the same region 
$IV$. Let us briefly discuss their specific features.

\begin{figure}
    \centering
    \includegraphics[width=1\linewidth]{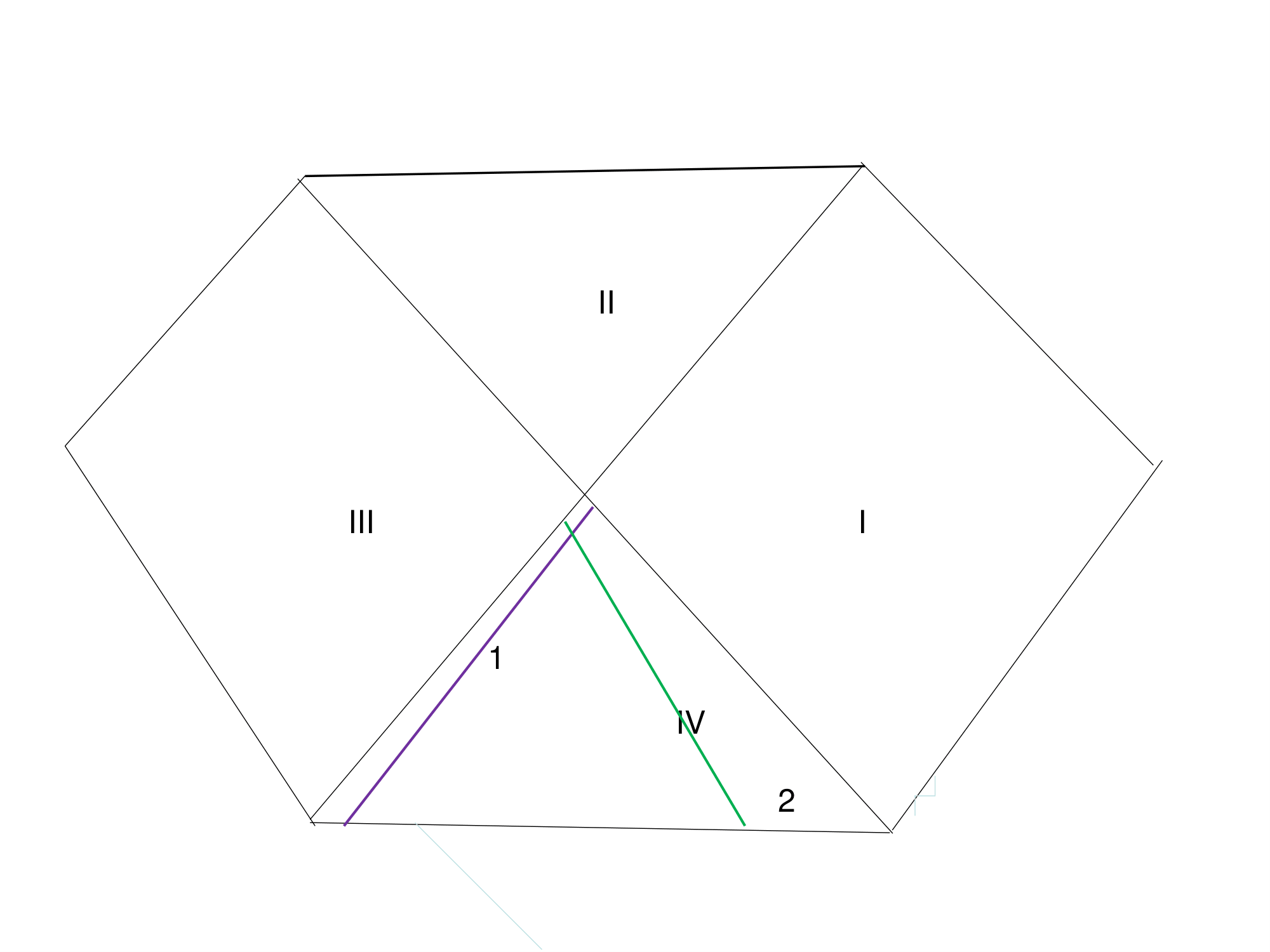}
    \caption{Scenario 7.}
    \label{7}
\end{figure}

\begin{figure}
    \centering
    \includegraphics[width=1\linewidth]{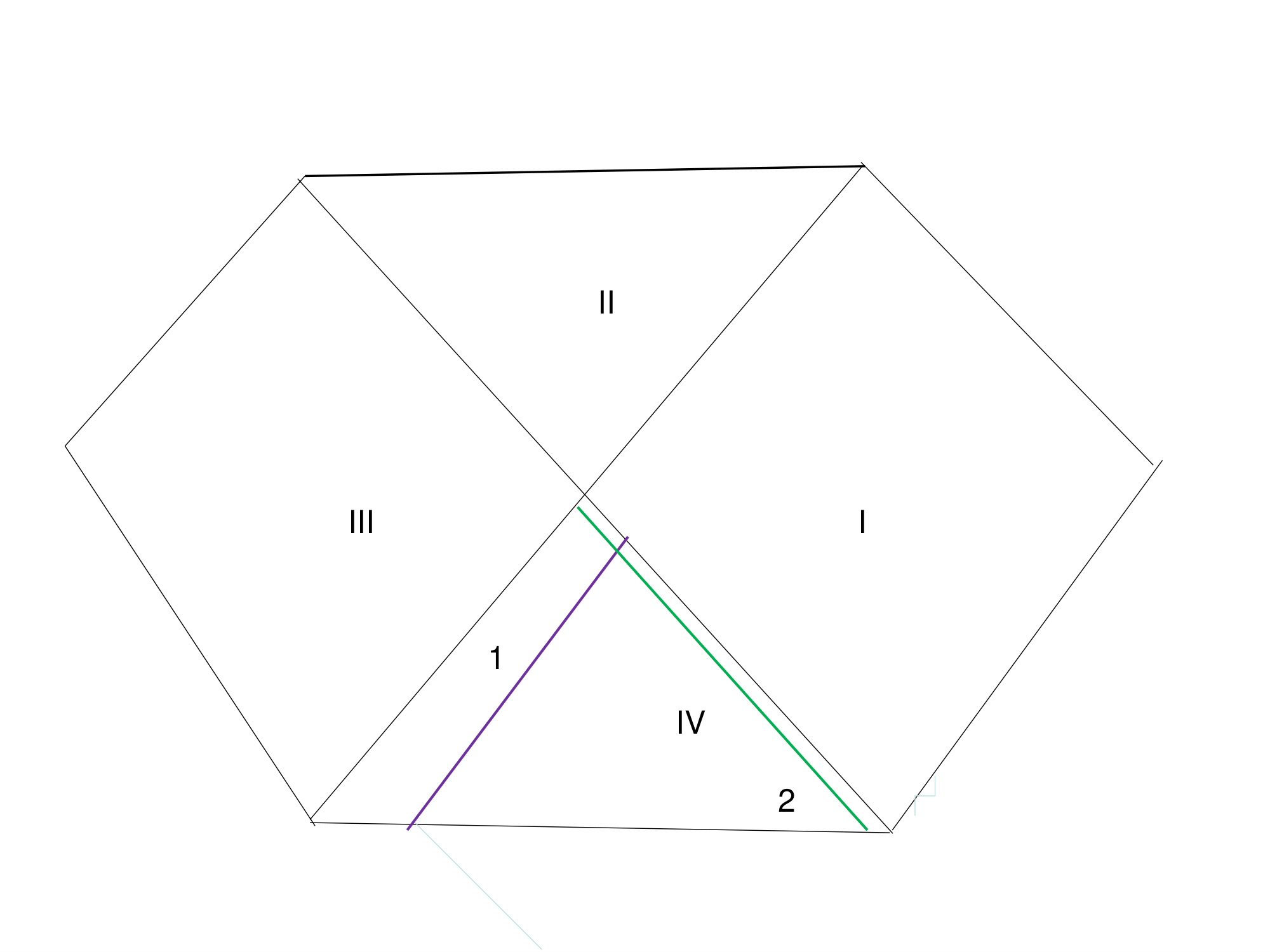}
    \caption{Scenario 8}
\end{figure}

We have from eq. (\ref{1}) with $\sigma ^{\prime }=1$ that%
\begin{equation}
\frac{dr}{dt}=-\sigma \left\vert f\right\vert \frac{\sqrt{\varepsilon
^{2}+\left\vert f\right\vert }}{\varepsilon }\text{,}
\end{equation}%
whence%
\begin{equation}
t=-\sigma \int_{0}^{r}\frac{\varepsilon dr^{\prime }}{\left\vert
f\right\vert \sqrt{\varepsilon ^{2}+\left\vert f\right\vert }}+t_{0}\text{,}
\end{equation}%
where $t_{0}=t(0)$.

It follows from Sec. \ref{kruskal} that the \ sign of $u^{X}$ coincides with
that of $\sigma $. If $\sigma =+1$ we attach label 1 to a particle. If $%
\sigma =-1$ we attach label 2 to a particle.

For both scenarios 7 and 8, for particle 1 $\sigma =1$ and%
\begin{equation}
t_{1}(r)=-\int_{0}^{r}\frac{dr^{\prime }\varepsilon _{1}}{\left\vert
f\right\vert \sqrt{\varepsilon _{1}^{2}+\left\vert f\right\vert }}%
+t_{0}^{(1)}\text{.}  \label{t1}
\end{equation}

For particle 2 $\sigma =-1$ and%
\begin{equation}
t_{2}(r)=\int_{0}^{r}\frac{dr^{\prime }\varepsilon _{2}}{\left\vert
f\right\vert \sqrt{\varepsilon _{2}^{2}+\left\vert f\right\vert }}%
+t_{0}^{(2)}\text{.}  \label{t2}
\end{equation}

For collision to occur, we require $t_{1}(r_{c})=t_{2}(r_{c})$, hence%
\begin{equation}
t_{0}^{(2)}-t_{0}^{(1)}=-(\int_{0}^{r_{c}}\frac{dr^{\prime }\varepsilon _{1}%
}{\left\vert f\right\vert \sqrt{\varepsilon _{1}^{2}+\left\vert f\right\vert 
}}+\int_{0}^{r_{c}}\frac{dr^{\prime }\varepsilon _{2}}{\left\vert
f\right\vert \sqrt{\varepsilon _{2}^{2}+\left\vert f\right\vert }})\text{.}
\label{0t21}
\end{equation}

Our goal is to arrange collision near the horizon, so $r_{c}\rightarrow
r_{+} $. Then, the main contribution in the integral comes from the vicinity
of the horizon, where $\left\vert f\right\vert \approx \frac{r_{+}-r}{r_{+}}$%
, so%
\begin{equation}
\int_{0}^{r_{c}}\frac{dr^{\prime }\varepsilon }{\left\vert f\right\vert 
\sqrt{\varepsilon ^{2}+\left\vert f\right\vert }}\approx -r_{+}\ln
\left\vert \frac{r_{+}-r_{c}}{r_{+}}\right\vert +a\text{,}
\end{equation}%
where $a$ is some constant, and%
\begin{equation}
t_{1}=t_{2}=t_{c}\approx r_{+}\ln \left\vert \frac{r_{+}-r_{c}}{r_{+}}%
\right\vert +t_{0}^{(1)}-a,
\end{equation}%
\begin{equation}
t_{0}^{(2)}-t_{0}^{(1)}\approx 2r_{+}\ln \left\vert \frac{r_{+}-r_{c}}{r_{+}}%
\right\vert .  \label{t12}
\end{equation}

Now, we consider subcases.

\subsubsection{Collision near the left horizon (Scenario 7)}

In terms of Kruskal-Szekeres coordinates,

\begin{equation}
U=-\sqrt{1-\frac{r}{r_{+}}}\exp (\frac{r}{2r_{+}})\exp (-\frac{t}{2r_{+}})%
\text{,}
\end{equation}%
\begin{equation}
V=-\sqrt{1-\frac{r}{r_{+}}}\exp (\frac{r}{2r_{+}})\exp (\frac{t}{2r_{+}}).
\end{equation}%
We require We require%
\begin{equation}
U_{c}\approx 0\text{, }V_{c}=O(1)\text{.}  \label{left}
\end{equation}

For particle 1%
\begin{equation}
U_{c}^{(1)}\approx -\sqrt{e}\exp (-\frac{t_{0}^{(1)}}{2r_{+}}+\frac{a}{2r_{+}%
})  \label{U7}
\end{equation}%
\begin{equation}
V_{c}^{(1)}\approx \sqrt{e}\exp (\frac{t_{0}^{(1)}}{2r_{+}}-\frac{a}{2r_{+}}%
)(1-\frac{r_{c}}{r_{+}})  \label{V7}
\end{equation}

To obey (\ref{left}), we choose%
\begin{equation}
t_{0}^{(1)}\approx -2r_{+}\ln (1-\frac{r_{c}}{r_{+}})+const\rightarrow
+\infty \text{,}
\end{equation}

\begin{equation}
t_{0}^{(2)}\text{is finite}
\end{equation}

Thus particle 1 starts from the far right end of a leg of a cylinder in the
singular state. Then, 
\begin{equation}
t_{1}\approx -r_{+}\ln (1-\frac{r_{c}}{r_{+}})\rightarrow +\infty .
\end{equation}

\subsubsection{Collision near the right horizon (Scenario 8)}

For particle 1 eqs. (\ref{t1}), (\ref{t2}), (\ref{U7}), (\ref{V7}), (\ref%
{0t21}) are still valid but the requirement are different. Now, we want to
have%
\begin{equation}
U_{c}=O(1)\text{, }V_{c}\approx 0.
\end{equation}

To this end, we choose $t_{0}^{(1)}$ to be finite, so%
\begin{equation}
t_{0}^{(2)}\approx 2r_{+}\ln \left\vert \frac{r_{+}-r_{c}}{r_{+}}\right\vert
\end{equation}%
and 
\begin{equation}
t_{c}\approx r_{+}\ln (1-\frac{r_{c}}{r_{+}})\rightarrow -\infty \text{.}
\end{equation}

Then,%
\begin{equation}
t_{1}=t_{2}\approx r_{+}\ln (1-\frac{r_{+}}{r})+t_{0}^{(1)}\rightarrow
-\infty
\end{equation}%
\begin{equation}
t_{0}^{(2)}\approx 2r_{+}\ln \left\vert \frac{r_{+}-r_{c}}{r_{+}}\right\vert
\rightarrow -\infty
\end{equation}

Eqs. (\ref{t2}), (\ref{0t21})%
\begin{equation}
t_{2}=\int_{0}^{r}\frac{dr^{\prime }}{\left\vert f\right\vert \sqrt{%
\varepsilon ^{2}+\left\vert f\right\vert }}+t_{0}^{(2)}
\end{equation}%
\begin{equation}
t_{0}^{(2)}-t_{0}^{(1)}=-2\int_{0}^{r_{c}}\frac{dr^{\prime }}{\left\vert
f\right\vert \sqrt{\varepsilon ^{2}+\left\vert f\right\vert }}
\end{equation}

We can compare both scenarios in a Table.

\begin{tabular}{|l|l|l|}
\hline
& Scenario 7 & Scenario 8 \\ \hline
$\sigma _{1}$ & $1$ & $1$ \\ \hline
$\sigma _{2}$ & $-1$ & $-1$ \\ \hline
$t_{c}$ & $\infty $ & $-\infty $ \\ \hline
$t_{0}^{(1)}$ & $+\infty $ & finite \\ \hline
$t_{0}^{(2)}$ & finite & $-\infty $ \\ \hline
\end{tabular}

\bigskip

\section{Properties of scenarios in terms of the Lema\^{\i}tre time\label%
{lem}}

In the previous Sections, we have considered possible high-energy collisions
using Kruskal-Szekeres coordinates. In order to give a better physical
interpretation, we would like to deal also with a physically relevant
observer that is able to probe a region inside a black (white) hole. To this
end, we fix a geodesically moving observer which can detect the collisions
in question. This needs to fix the corresponding free falling frame which
cannot cover the whole space-time. There exist four different Lema\^{\i}%
tre-like systems living in four different domains, so that a domain consists
of two space-time zones of the full geometry. There are $(R+,T-)$ pair which
involve our region $R+$ and a black hole region $T-$, $(T+,R+)$ which
include a white hole and our region, and the combination $(R-,T-)$ and $%
(T+,R-)$ which are combinations of a black and white hole and the "mirror"
world $R-$ correspondingly. Since in each case below we indicate what a
particular frame we deal with, we simply use the single symbol $\tilde{t}$
to denote the Lema\^{\i}tre time and do not use subscript (strictly
speaking, there are four different times $\tilde{t}$). (For more details
about connection between a space-time region and type of the Lema\^{\i}tre
time see \cite{genrad}.)

Let us start with scenarios 1 and 2. (Hereafter, we use for shortness
expressions "far past" or "far future" that simply means that the Lema\^{\i}%
tre time is large negative or large positive.) Suppose, our observer lives
in a white hole and correspondingly uses the expanding Lema\^{\i}tre frame
with time $\tilde{t}$. This time coordinate is regular at the horizon
between $I$ and $IV$ and diverges at the boundary of the domain located at
the horizon between $I$ and $II$. As for the free-falling time from (to) the
horizon $(I-IV)$, it is finite for a particle comoving with the frame (going
from the horizon outward in our case), and diverges for the "wrong"
direction when the horizon is being approached.

The 1-st scenario \textit{from the viewpoint of a white hole observer} can
be interpreted as follows. Particle 1 appears from a singularity at some
finite $\tilde{t}_{1}(1)_{0}$ and crosses the horizon at some another finite 
$\tilde{t}(1)_{1}$. Particle 2 appears in the $R+$ region close to the
horizon at the diagram, which means that its initial $\tilde{t}(2)_{0}$ is
big and negative. So that, the 2-nd particle emerges from an infinity in
infinite past of an expanding Lema\^{\i}tre frame, moves \textit{towards}
the black hole (where it would find itself without collision), and meets
particle 1 at some finite $\tilde{t}$. The infinity in $\tilde{t}(2)_{0}$ is
compensated by an infinite travel time towards the horizon (since particle 2
moves in the "wrong" direction in terms of expanding frame), and the
collision occurs at a finite Lema\^{\i}tre time. The actual value of $\tilde{%
t}$ is not important for the energy of collision, it is $\tilde{t}(2)_{0}$
which matters, and more negative $\tilde{t}(2)_{0}$ leads to bigger
collision energy.

As for the scenario 2, particle 1 starts its motion at some finite $\tilde{t}%
(1)_{0}$, moves further almost along the F-horizon, so the corresponding Lema%
\^{\i}tre time almost diverges. Meanwhile, particle 2 \ needs almost
diverging Lema\^{\i}tre time independently of its initial position at some $%
\tilde{t}(2)_{0}$. So that, the collision occurs for very big $\tilde{t}$.
Bigger $\tilde{t}$ leads to bigger collision energy since it points gets
closer to the horizon.

The white hole interpretation is the most natural one for scenarios 1 and 2.
However, suppose that we have a black hole instead. The contracting Lema%
\^{\i}tre frame exists in the domain $(R+,T-)$. \ The part of the trajectory
belonging to the $T+$ zone is inaccessible to a black hole observer, so the
trajectory of particle 1 starts from "nothing" (if an observer attached to
particle 2 tries to describe it). Physically, it may indicate that particle
1 has been born in some other physical process like scattering or decay
which is not consider here. From such a point of view these two scenarios
appear to be quite similar. The contracting Lema\^{\i}tre time at the
P-horizon diverges (it is a boundary of the black hole domain), the time
travel \thinspace $\tilde{t}$ of particle 1 from the horizon diverges as
well (since the outward direction is now the "wrong" one), these two
infinities compensate each other, and the collision with particle 2 occurs
at some finite $\tilde{t}$, where $\tilde{t}$ is the \textit{contracting}
Lema\^{\i}tre time now).

Interpretation of the scenarios 5 and 6 is the same, with the replacement of
"our" asymptotically flat zone to the "mirror" one. We can use either mirror
expanding or mirror contracting Lema\^{\i}tre frames.

Trajectories in scenarios 3 and 4 go through both $R+$ and $R-$ zones.
Collisions occur in the $T-$ zone. Suppose, an observer belongs to our
world. Then, two scenarios describe qualitatively similar physical
situations. We remind a reader that the free falling time to/from a horizon
in the $T+$ or $T-$ region is finite for the correct sign of $\sigma $ (this
sign is positive for the contracting frame in "our" world) and infinite for
the wrong sign (which is negative for "our" frame in the case under
discussion). So that, particle 1 emerges from the mirror Universe where the
frame with contracting Lema\^{\i}tre time does not exist at all. Then, it
crosses the horizon and appears in region $II$. It has $\sigma <0,$ so the
Lema\^{\i}tre time between any point close to the horizon and some other
point with finite $r<r_{+}$ would diverge. However, as the point of
collision itself is close to $r_{+}$, the limiting procedure (see below) can
give a finite $\tilde{t}$. Then, it collides with an usual ($\sigma >0$)
particle falling into a black hole from outside, also at some finite $\tilde{%
t}$.

Finally, scenarios 7 and 8 involve $T+$ region, i.e. a white hole. Let us
suppose that we consider the process in the expanding Lema\^{\i}tre frame
that covers the $T+$ region and our $R+$ one. Then, in scenario 8 particle 1
emerges from a singularity at finite expanding Lema\^{\i}tre frame time with 
$\sigma >0$ while particle 2 emerges from a singularity in far past with $%
\sigma <0$. The particle 2 needs infinite $\tilde{t}$ to reach the horizon
between regions $I$ and $IV$, so two infinities cancels out, and it collides
with particle 1 at finite $\tilde{t}$. In scenario 7 particle 2 emerges at a
finite $\tilde{t}$, so particle 1 should start at far future in order to
meet particle 2. For the "mirror" expanding frame that connects a white hole
and the mirror Universe the roles of particles are reversed.

We see that in scenarios 1, 2, 5, 6, 7 and 8 a white hole is an essential
ingredient. Either at least one of particles passes through it or collision
occurs there.

Each of scenarios requires some kinematic restrictions for its realization.
For some particular scenarios, these conditions were considered in \cite%
{white-black} using the Kruskal-Szekeres coordinates or even standard
Schwarzschild-like ones. See also next Section. Meanwhile, it is instructive
to give also the corresponding picture directly in terms of the Lema\^{\i}%
tre time. Below, we consider as an example scenario 3.

Let us, for definiteness, use the standard Lema\^{\i}tre time $\tilde{t}$
related to a free falling observer \cite{LL}. Then, for a part of trajectory
of particle 1 between some point $r_{1}$ near the horizon and the point of
collision $r_{c}$ we have%
\begin{equation}
\tilde{t}=\int_{r_{c}}^{r_{1}}\frac{dr}{P\left\vert f\right\vert }%
(\varepsilon +P_{0}P)+\tilde{t}_{0}
\end{equation}%
where $\tilde{t}_{0}$ is a constant,%
\begin{equation}
P_{0}=\sqrt{\varepsilon _{1}^{2}+\left\vert f\right\vert }\text{,}
\end{equation}%
\begin{equation}
P=\sqrt{1+\left\vert f\right\vert },
\end{equation}%
more general case is discussed in Sec. 2 of Ref. \cite{genrad}.

Let $r_{1}=r_{+}(1-\delta )$, $r_{c}=r_{+}(1-\alpha )$, where $0<\delta \ll
1 $, $0<\alpha \ll 1$ with $\alpha >\delta $ since $r$ is decreasing. Then,%
\begin{equation}
\tilde{t}\approx 2\varepsilon _{1}\ln \frac{\delta }{\alpha }+\tilde{t}_{0}%
\text{.}
\end{equation}%
In the point of collision the Lema\^{\i}tre time $\tilde{t}$ of both
particles coincides, $\tilde{t}_{1}=\tilde{t}_{2}$, so we must choose 
\begin{equation}
\tilde{t}_{0}=\tilde{t}_{1}+2\varepsilon _{1}\ln \frac{\alpha }{\delta }+C%
\text{,}
\end{equation}%
where $C$ is another unessential finite constant. We see that if $\alpha $
and $\delta $ have the same order, $\tilde{t}_{0}$ is finite. If $\delta \ll
\alpha $, it becomes indefinitely large.

\section{Collisions of massive particles with resting observers and
bifurcation point\label{mar}}

So far, we have seen that infinite energy collisions at the horizon require
either different direction of motion of two particles (in a $R$ region) or
different signs of their $\sigma $ (in a $T$ region). However, one more case
is still beyond our considerations - if $\sigma =0$. Such a trajectory is a
geodesic which can exist only in $T$ region \cite{dor}. It is this observer
that is called a resting one $(RO)$. Equivalently, one can put $\varepsilon
=0$.

In this and the following sections we are going to take into account
scenarios in which such a kind of observers participates. Let us consider
scenario in which particle 1 passes through the bifurcation point, $%
\varepsilon _{1}=0$ and $y=y_{1}=const$. In this section, we will assume
that both particles are massive.

The picture describing such scenarios is given below. There are two of them.

Scenario BIF1. Particle 2 (which is not $RO$) comes from our Universe.

\begin{figure}
    \centering
    \includegraphics[width=1\linewidth]{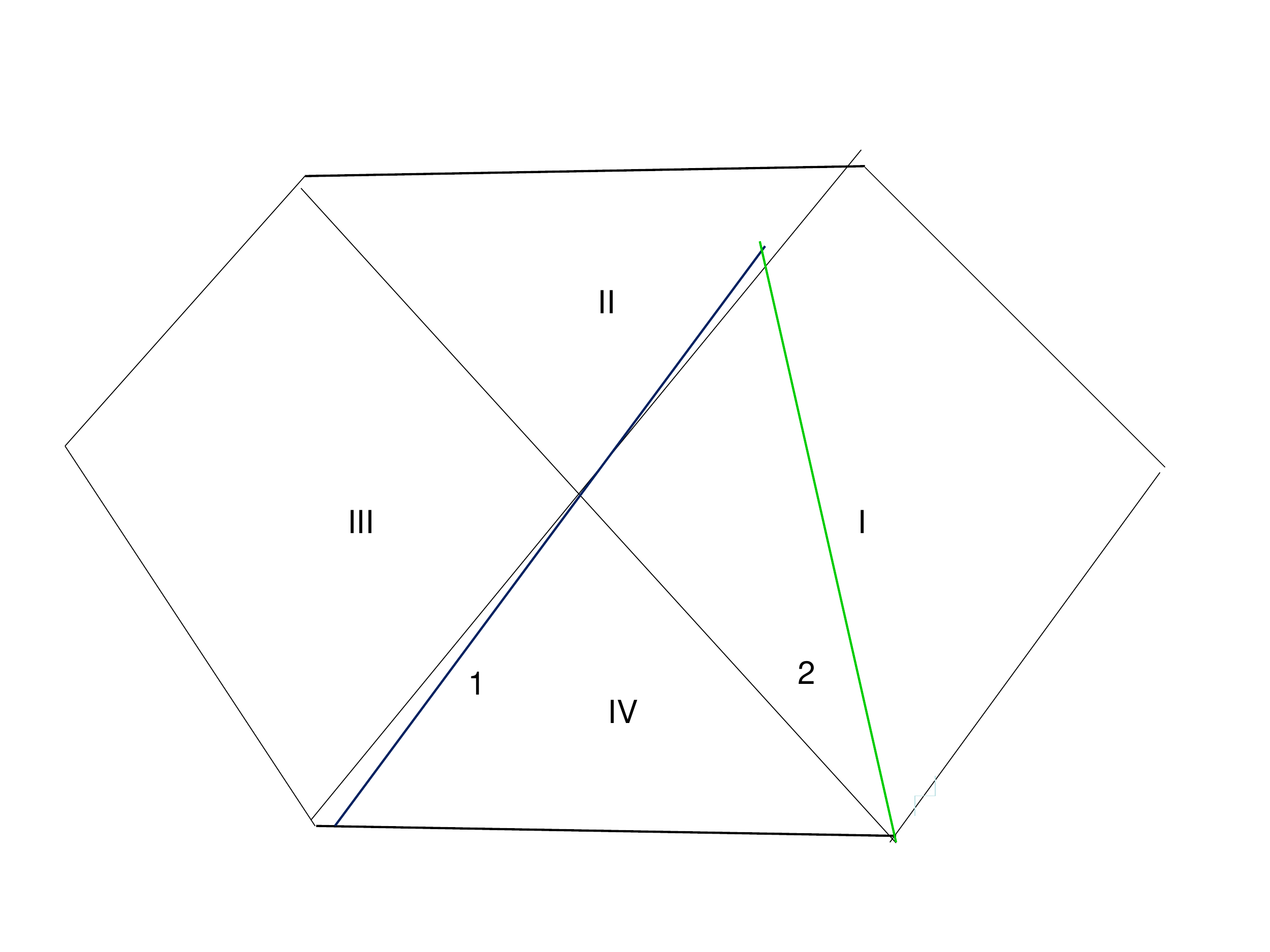}
    \caption{Scenario BIF1. Particle 1 is RO. Particle 2 comes from our Universe.}
\end{figure}

It is worth noting that the point of collision can be at any distance from
the bifurcation point, $V_{c}$ is arbitrary. Hence, by continuous
deformation it can be modified in such a way that collision would occur with
small $V_{c}$, thus close to the bifurcation point. However, since the
bifurcation point itself is inaccessible for the usual particle (not $RO$),
this would not form a new scenario.

It follows from (\ref{ga}) that now

\begin{equation}
\gamma =\frac{\sqrt{\varepsilon _{2}^{2}+\left\vert f\right\vert }}{\sqrt{%
\left\vert f\right\vert }}.
\end{equation}%
For $\left\vert f\right\vert \rightarrow 0$, we have 
\begin{equation}
\gamma \approx \frac{\varepsilon _{2}}{\sqrt{\left\vert f\right\vert }}%
\rightarrow \infty
\end{equation}%
that corresponds to eq. (10.2) of \cite{jcap}. We can see that in this case $%
\gamma $ grows as $\left\vert f\right\vert ^{-1/2}$ i.e. slower than in the
case of head-on-collisions, where $\gamma \sim \left\vert f\right\vert ^{-1}$%
.

If both particles are $RO,$ $\varepsilon _{1}=\varepsilon _{2}=0$, so $%
\gamma =1$ and high energy collision is impossible. Moreover, the fact that $%
\gamma =1$ means that any two $RO$ have zero relative velocity with respect
to each other.

Collisions of such a type were considered in \cite{jcap} but the analysis
remained incomplete since the kinematic conditions necessary for collision
were not investigated there. Below, we fill this gap. Now, we are interested
in the kinematic conditions which make it possible for two particles to
collide near the horizon, with $\left\vert f\right\vert \ll 1$.

1) Scenario BIF1. We assume that there are no restrictions on particle 2.
Collision is supposed to occur near the future horizon, so $U_{c}\approx 0$, 
$V_{c}=O(1)$.

For the trajectory of particle 1 in the black hole region,%
\begin{equation}
U=\exp (-\kappa u)\text{, }V=\exp (\kappa v)\text{,}  \label{U}
\end{equation}%
\begin{equation}
u=t-r^{\ast }\text{, }v=t+r^{\ast }\text{,}
\end{equation}%
\begin{equation}
r^{\ast }=r+r_{+}\ln \frac{r_{+}-r}{r_{+}}\text{,}
\end{equation}%
$\kappa $ is the surface gravity. For the Schwarzschild metric $\kappa =%
\frac{1}{2r_{+}}$. Then,%
\begin{equation}
\frac{U}{V}=\exp (-2\kappa t)=\exp (-2\kappa y).  \label{uvt}
\end{equation}

For it, $t=y=y_{0}=const$. Now, $U_{c}\ll V_{c}$. Thus we must choose 
\begin{equation}
y_{0}>0,\kappa \left\vert y_{0}\right\vert \gg 1  \label{yt}
\end{equation}%
.

Eq. (\ref{U}) is a coordinate transformation. However, one can check by
direct calculations from geodesic equations of motion for such a particle, $%
\varepsilon =0$ that the relation (\ref{uvt}) follows again. Thus the
condition (\ref{yt}) is sufficient and necessary for realization of the
trajectory within the scenario under discussion. There are no special
restrictions for the trajectory of particle 2.

2) Scenario BIF2. Particle 2 (which is not $RO$) comes from the mirror
Universe.

\begin{figure}
    \centering
    \includegraphics[width=1\linewidth]{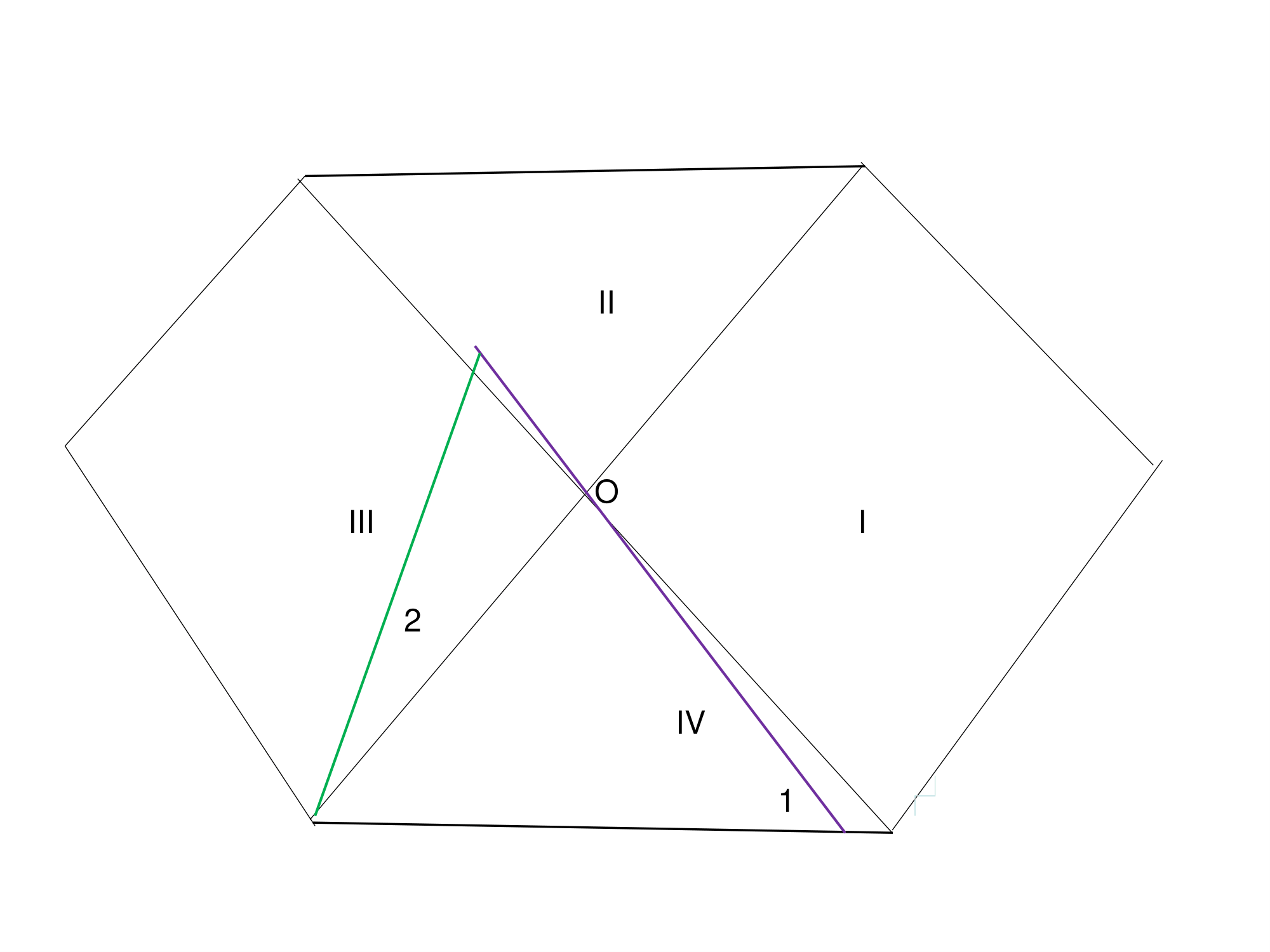}
    \caption{Scenario BIF 2. Particle 1 is RO. Particle 2 comes from mirror Universe.}
\end{figure}

Thus, we have two distinct scenarios for high energy collisions in the
region $T-$ (region II) in which $RO$ participates\textbf{. }

Similar options exist in the $T+$ (region IV). Now, before a $RO$ would pass
trough the bifurcation point, it experiences collision with a "usual"
particle (not $RO$). See the plots corresponding to scenarios BIF3 and BIF4.

\begin{figure}
    \centering
    \includegraphics[width=1\linewidth]{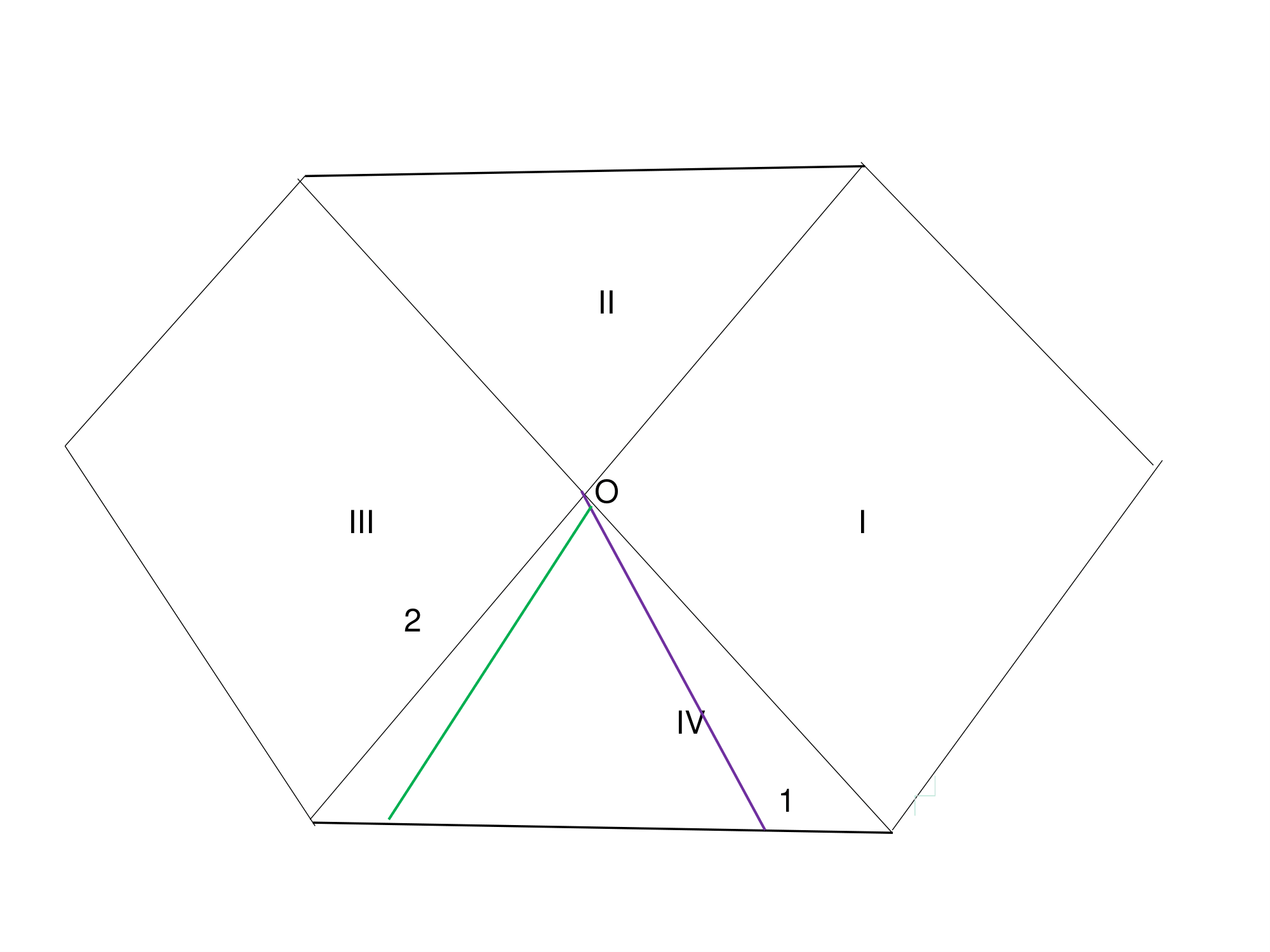}
    \caption{Scenario BIF3. Particle 1 is RO.}
\end{figure}

\begin{figure}
    \centering
    \includegraphics[width=1\linewidth]{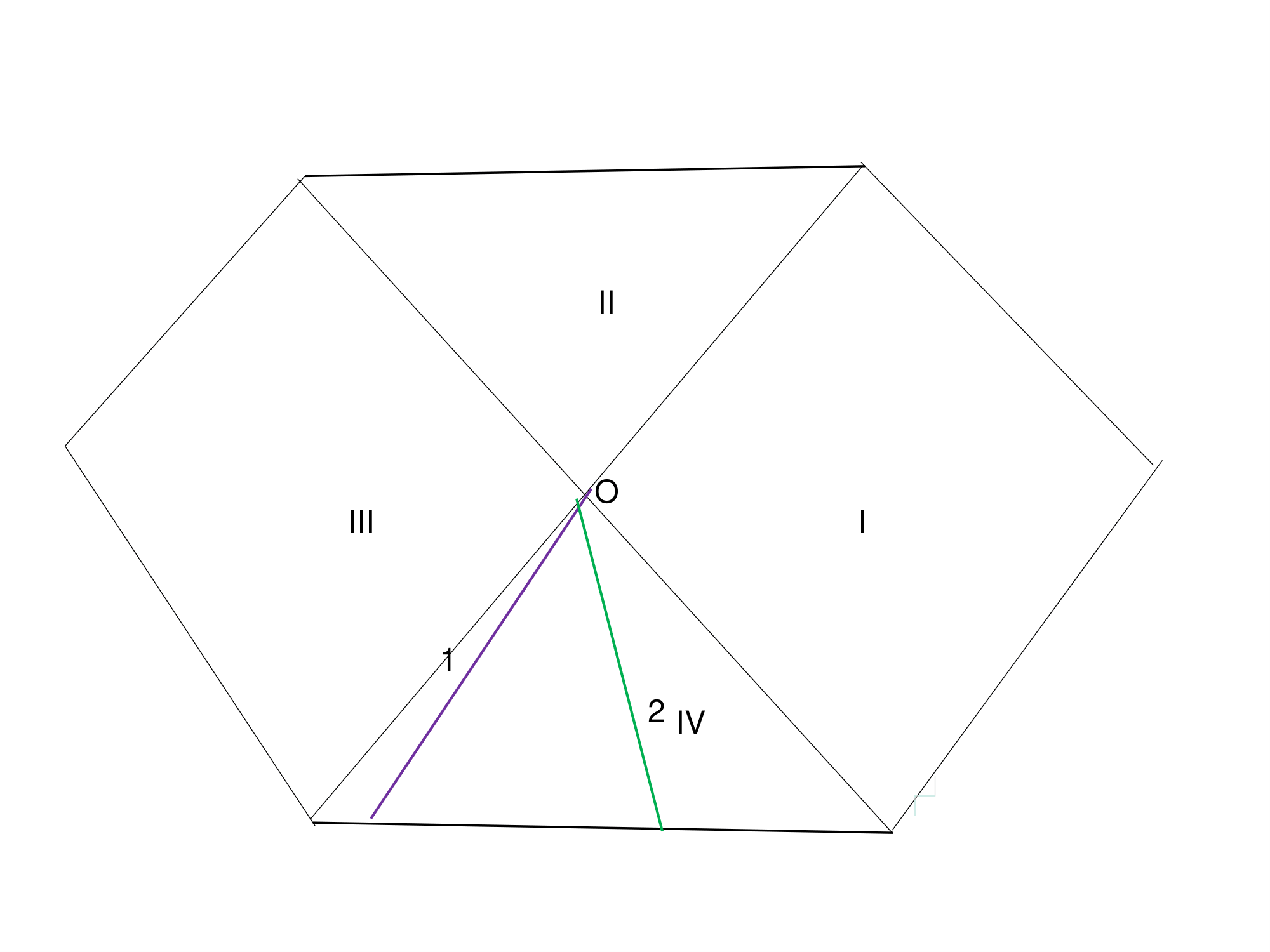}
    \caption{Scenario BIF4. Particle 1 is RO.}
\end{figure}

The interpretation of these scenarios from the viewpoints of Lema\^{\i}tre
time is based on the properties of the corresponding frames described in the
beginning of Sec. \ref{mar}. There are four different frames. Although each
of these frame is incomplete, their advantage consists in that they are
formed by free falling observers and in this sense they are physical. (See
more on this in \cite{genrad}.). As observers of this type are unable to see
the whole picture, they perceive that some particles which they are watching
appear "from nothing". Let us consider the scenarios under discussion in
this context.

Collisions in the scenarios BIF1 and BIF2 occur in the $T_{-}$ region, so
only observers of the types $(R_{-},T_{-})$ and $(R_{+},T_{-})$ can witness
them. For any such an observer the 1-st particle (with $\varepsilon =0$)
appears from "nothing". In BIF1 an observer of type $(R_{-},T_{-})$ would
say that particle 2 appears from nothing as well. In BIF2 a similar
conclusion would be made by an observer of type $(R_{+},T_{-})$. For two
other cases an observer can see the particle 2 falling into $T_{-}$ region
from corresponding asymptotically flat region. In BIF3 and BIF4 the
collision is invisible for observers considered above. Only $(R_{-},T_{+})$
and $(R_{+},T_{+})$ observers can witness the collision. Exact details of
the scenarios can be easily reconstructed using the considerations of Sec. %
\ref{mar}, we do not stop here for this.


\section{Collisions of the massive particles with $\protect\varepsilon %
_{1,2}=O(1)$ near the bifurcation point\label{enear}}

One can argue that the requirement $\varepsilon =0$ exactly is not physical
since we can not fix a physical parameter with infinite precision, so that
actual $\varepsilon _{1}$ of the 1-st particle can be very small with $%
\sigma _{1}$ either positive or negative, falling (if $\sigma _{2}>0$) into
the description of the Sections \ref{sce}, \ref{lem} for a negative $\sigma
_{1}$ and giving only finite collision energy for a positive $\sigma _{1}$.

Indeed, we can consider a class of collisions where none of particles passes
through the bifurcation point but collision occurs close to it, $\left\vert
f\right\vert \ll 1$. Clearly, such scenarios can be obtained by continuous
deformation of the scenarios from the first group (where the particles
collide near an arbitrary point of the horizon - see Sec. \ref{sce}). (For
instance, in this manner, Scenario 3 from Ref. \cite{white-black} can be
obtained by the limiting transition from Scenario 1 in our Table I above.)
In doing so, the point of collision approaches the bifurcation one. Let us
illustrate this with the following example.

Let scenario in T+ region be realized when $\varepsilon _{1}\neq 0$, $%
\varepsilon _{2}\neq 0$, so both observers are not $RO$, but collision
occurs near the bifurcation point.

This means that in (\ref{U7}), (\ref{V7}) $U\rightarrow 0$, $V\rightarrow 0$%
, so $t_{0}^{(1)}\rightarrow +\infty $ and $\frac{t_{0}^{(1)}}{r_{+}}+\ln (1-%
\frac{r}{r_{+}})\rightarrow -\infty $

Correspondingly, it follows from (\ref{t12}) that%
\begin{equation}
t_{0}^{(2)}\approx t_{0}^{(1)}+2r_{+}\ln \left\vert \frac{r_{+}-r_{c}}{r_{+}}%
\right\vert .
\end{equation}%
\begin{equation}
t_{0}^{(2)}\rightarrow -\infty
\end{equation}%
and $t_{c}$ may take an arbitrary value depending on which quantity
dominates in (\ref{t1}), (\ref{t2}) - an integral or a constant.

\begin{figure}
    \centering
    \includegraphics[width=1\linewidth]{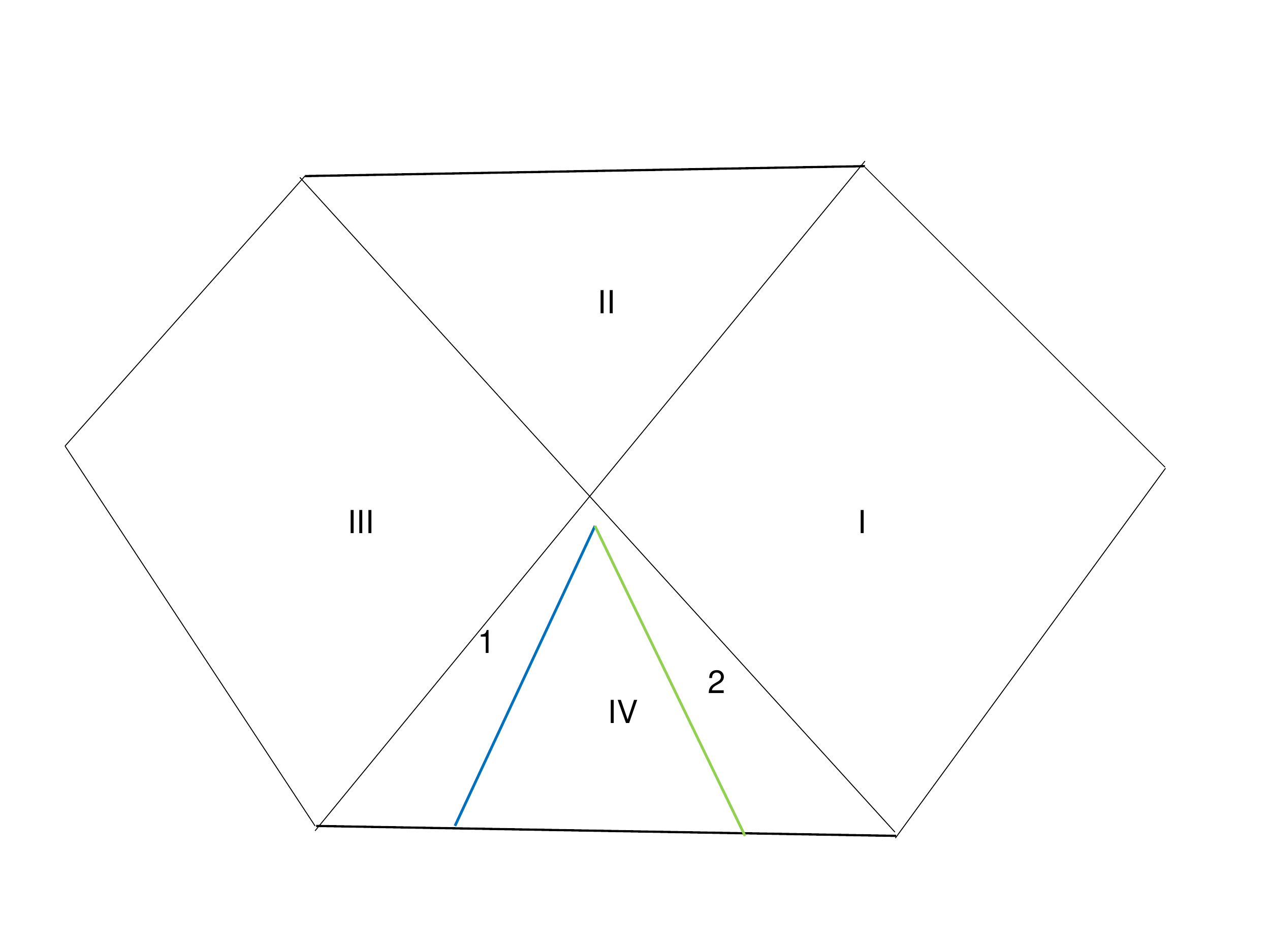}
    \caption{Collision near bifurcation point}
    \label{nearBFT}
\end{figure}

In the cases discussed in this Section $\varepsilon _{1,2}=O(1)$ the
standard dependence $1/\left\vert f\right\vert $ (\ref{gad}) is preserved.
This is in contrast with $1/\sqrt{\left\vert f\right\vert }$, the dependence
typical of the $RO$ presence, when $\varepsilon _{1}=0$ or $\varepsilon
_{2}=0$.

Meanwhile, it is worth noting that in general the actual dependence of $%
\gamma $ on $f$ is determined by the competition between $\varepsilon _{1}$, 
$\varepsilon _{2}$ in the numerator and $\left\vert f\right\vert $ in the
denominator. In principle, any intermediate case is possible depending on
the relation between $\varepsilon _{1},$ $\varepsilon _{2}$ and $\left\vert
f\right\vert .$

So that, in some cases it is reasonable to consider the situation with small
but nonzero $\varepsilon $ as a perturbed scenario of the previous section.
Consider, for example the scenario BIF1 where $\sigma _{2}>0$. Suppose that
actually $\sigma _{1}<0$ and its $\varepsilon _{1}$ is very small. Then,
remembering that a collision occurs not exactly at the horizon, but near the
horizon with a small but non-zero $f$, we see that if the collision in
question occurs when $\varepsilon _{1}\ll \sqrt{f}$, the dependence of the
collision energy on $f$ is still $\sim 1/\sqrt{f}$ as in the scenario BIF1
(10). Only for $f$ much less than $\varepsilon _{1}^{2}$ the dependence $%
\sim 1/f$ restores.

Similar way, in the case $f\gg \varepsilon _{1}^{2}$ we still have $\gamma
\sim 1/\sqrt{f}$ until $f$ becomes small enough and the final finite
asymptotic (\ref{gae}) is reached.


\section{Illusory scenarios: may the kinematic censorship be violated?\label%
{il}}

It is instructive to make a short comment on seeming violation of the
kinematic censorship. Let us consider, as an example, collision of particles
inside a white hole. Each of them can reach the white hole horizon for a
finite proper time. If they have $\sigma _{1}=-\sigma _{2}$, Eq. (\ref{gag})
it would seem that they can collide on the (say) right horizon to give not a
large but literally infinite $\gamma $.

However, the key issue is that particles with different $\sigma $ move
towards different branches of the horizon. Then, three different cases are
possible.

\begin{figure}
    \centering
    \includegraphics[width=1\linewidth]{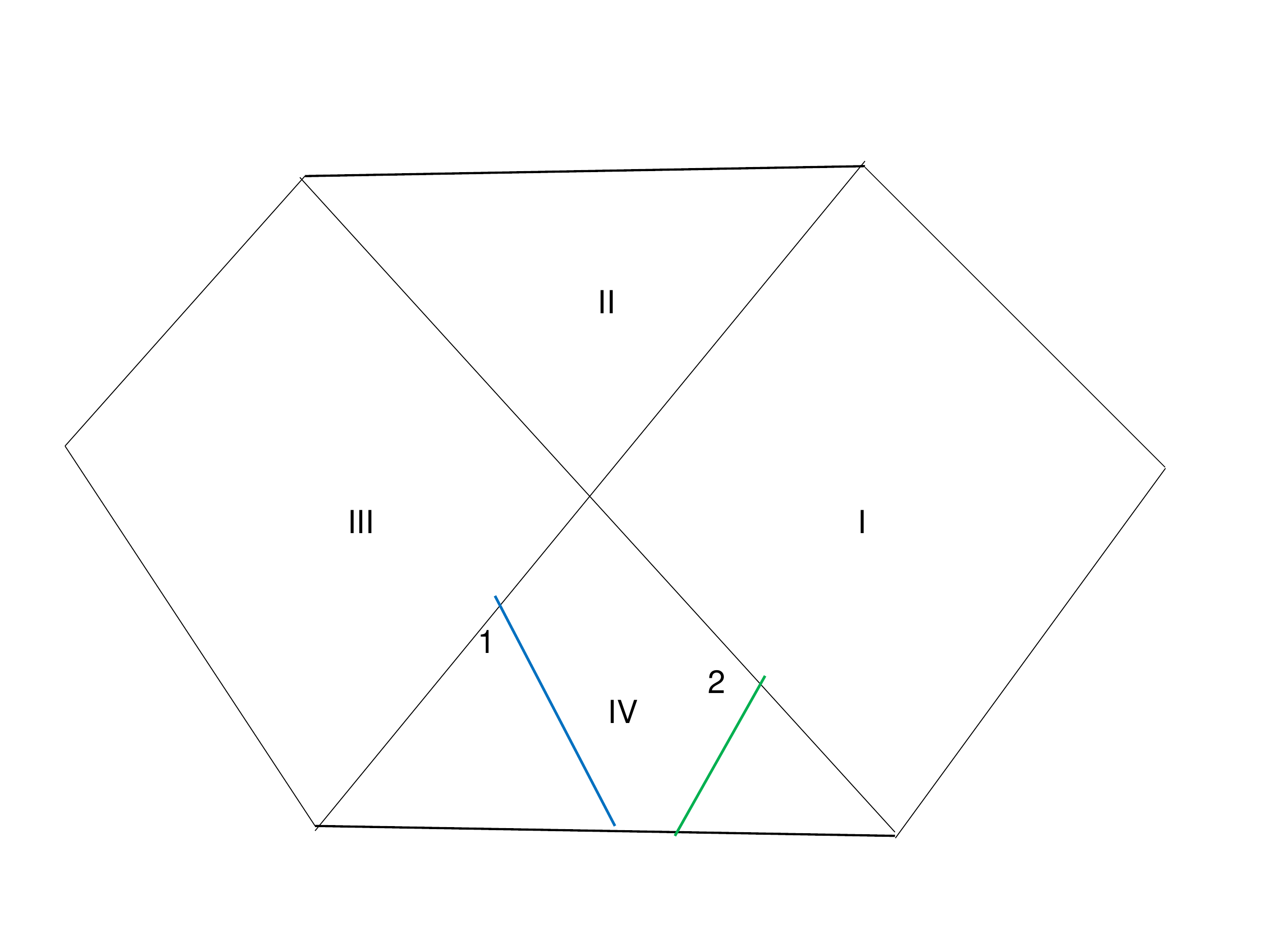}
    \caption{Collision on the horizon does not occur at all.}
    \label{16}
\end{figure}

(i) Collision on the horizon does not occur at all. See Fig. \ref{16}.

\begin{figure}
    \centering
    \includegraphics[width=1\linewidth]{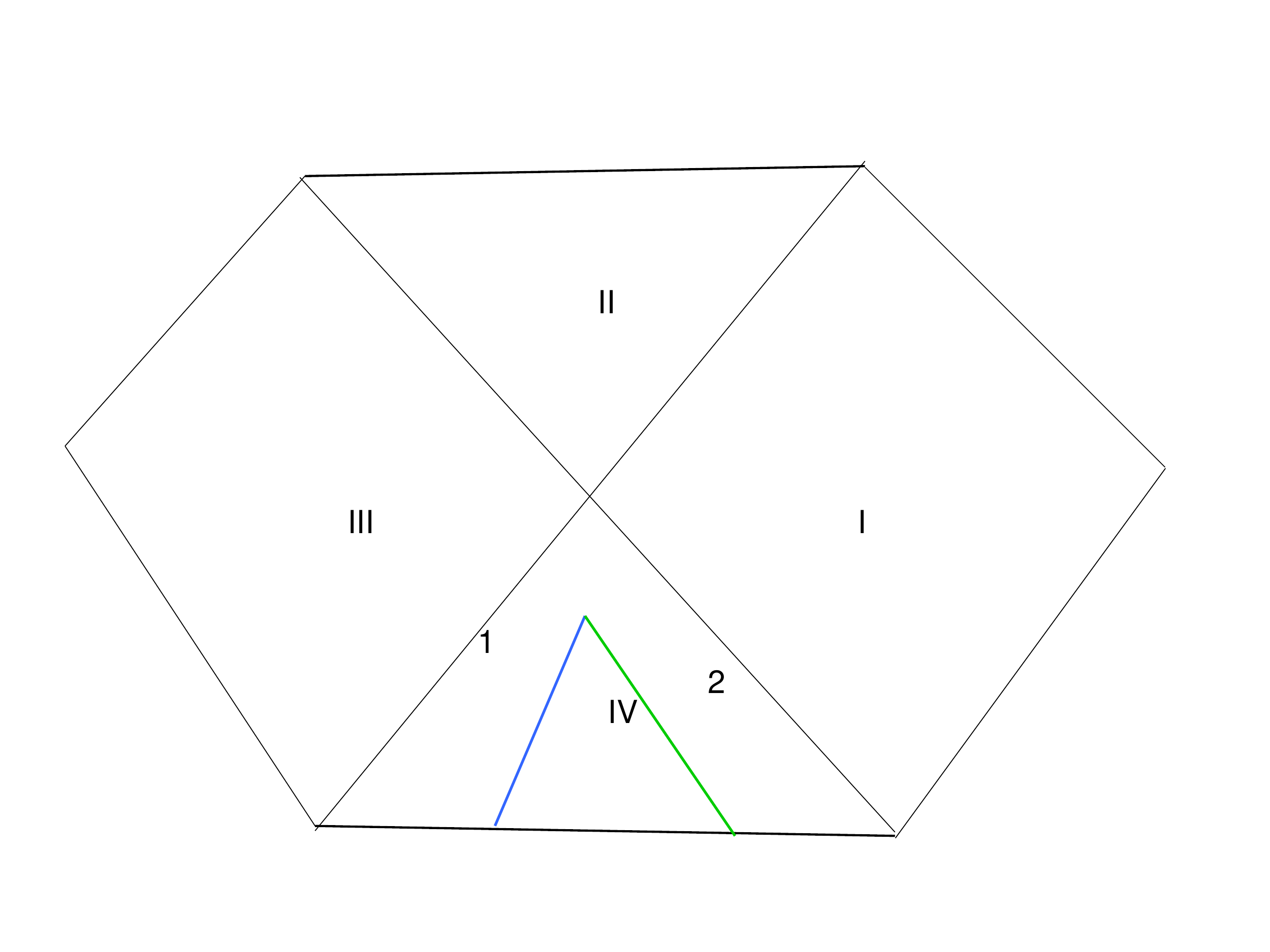}
    \caption{Collision happens at some intermediate point, hence with finite $\gamma$.}
    \label{inter}
\end{figure}

(ii) It can happen at some intermediate point, hence with finite $\gamma $.
See Fig. \ref{inter}.

(iii) It happens near the bifurcation point, so it is a particular case of
what has been considered above. See Fig. \ref{nearBFT}.

Hence, $\gamma $ can be large but cannot be infinite.

It is worth noting an additional formal option for getting infinite $\gamma $
that, however, cannot be realized. This would have occurred, had one of
particles moved exactly along the horizon to arrange collision exactly on
the horizon. But for a massive particle this is forbidden in principle.

\section{Collisions of the resting observers with massless particles near
the bifurcation point: classification of scenarios\label{massless}}

\begin{figure}
    \centering
    \includegraphics[width=1\linewidth]{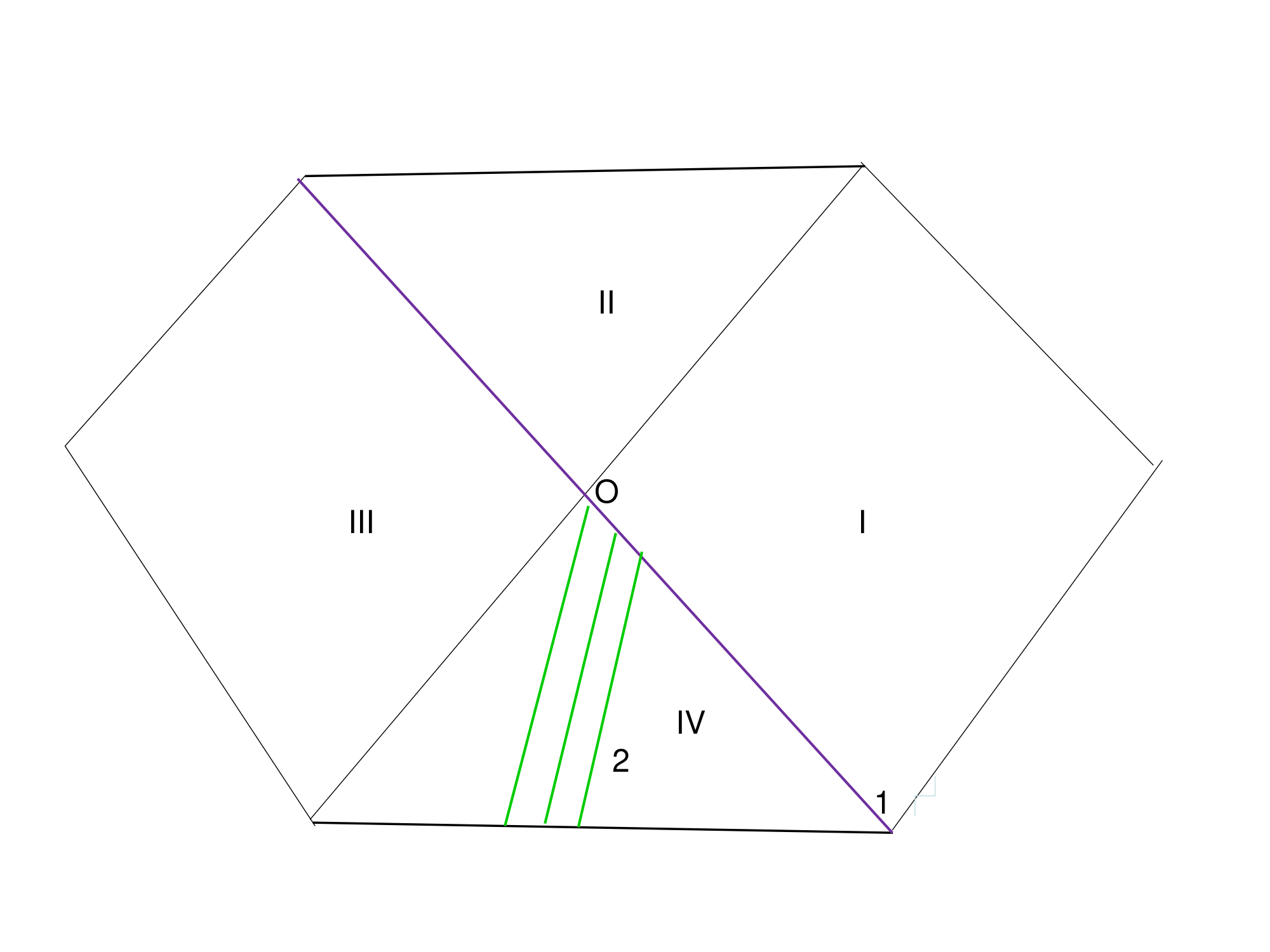}
    \caption{Scenario i. Photon 1 moves along horizon. Particle 2 is massive.}
    \label{Si}
\end{figure}

In the former Section, we assumed that $RO$ collides with another massive
particle. Meanwhile, a new perspective arises when the second particle is
massless (photon). This is because \ such a scenario includes collisions
with photons propagating along the horizon that has no analogue for a
massive particle. The situation when a massive particle (not $RO)$ collides
with \ a photon traveling along the horizon was considered in \cite{along}
(that generalized some previous results \cite{kas}). It turned out that in
the limit when the point of collision approaches the bifurcation point (BP),
blueshift grows unbounded. In Fig. \ref{Si} this happens when a point of
collision approaches closer and closer the BP. In the limit, our observer
becomes Resting Observer ($RO$). Then, a question arises: what happens in
the BP itself? If we simply take the limit in a straightforward manner, we
obtain infinity. It \textit{would seem} that such a $RO$ would perceive
those signals at BP as critically, i.e. infinitely, blueshifted. This would
break the "principle of the kinematic censorship" according to which the
energy released in any physical event cannot, literally, be infinite \cite%
{cens}. Such a paradox requires careful treatment of subtle details in the
limiting transitions. Let us consider what happens inside $WH$ region.

We will define the following three types of scenarios in which a massive
particle represents $RO$ or tends to it.

(i) Photon 1 moves along the horizon. Different time-like geodesics 2
originating within WH are crossing a horizon P (or F) closer and closer to
BP; the limiting transition occurs when such an observer turns into $RO$
that passes through the BP. See Fig. \ref{Si}.

\begin{figure}
    \centering
    \includegraphics[width=1\linewidth]{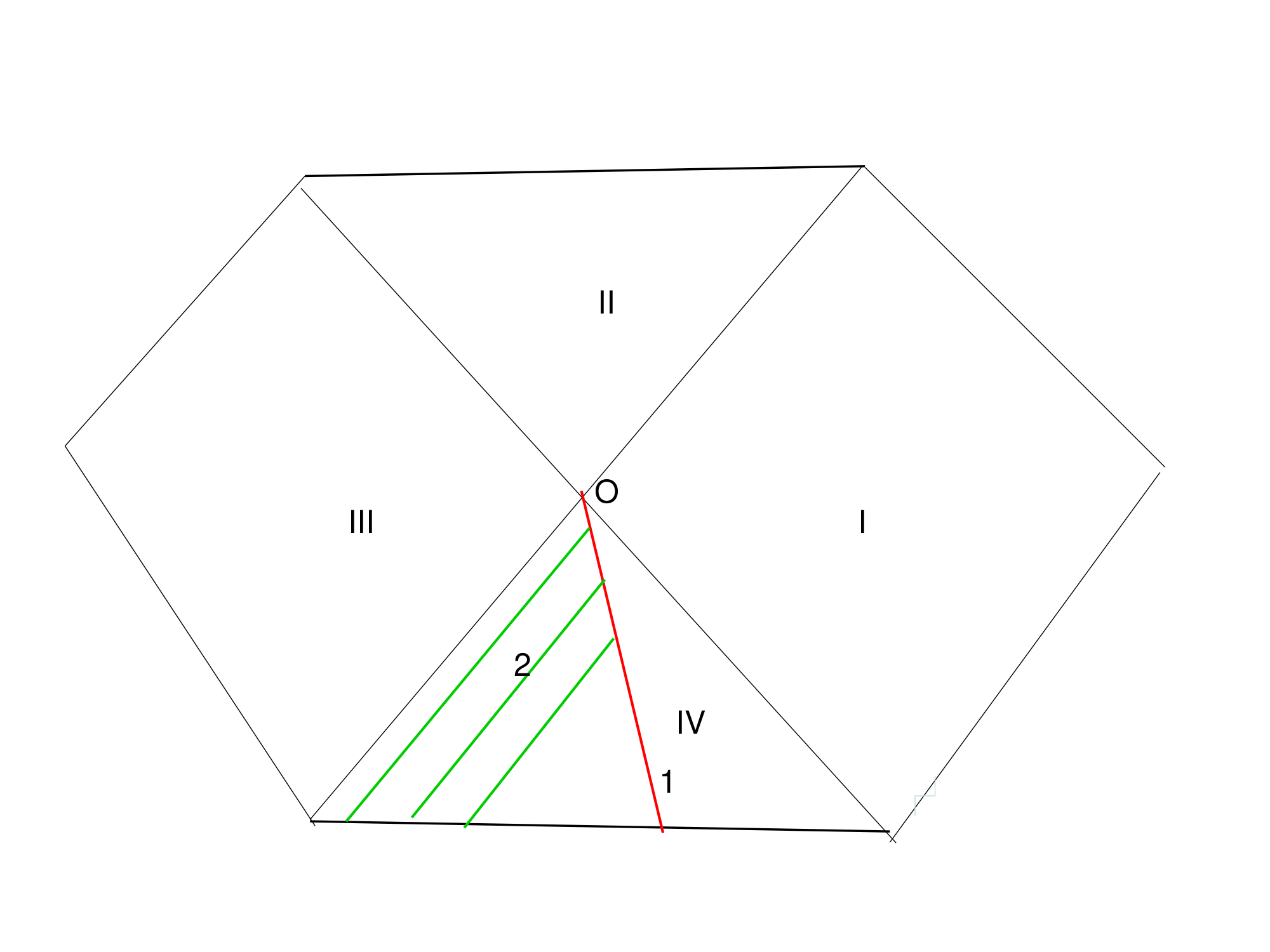}
    \caption{Scenario ii. Particle 1 is RO. Particle 2 is photon.}
    \label{ii-1}
\end{figure}

(ii) The scenario in which we fix a $RO$ (particle 1) from the very
beginning and consider photons 2 colliding with it closer and closer to the
horizon; in the limit the photon trajectory coincides with a photon
traveling along the horizon. See Fig. \ref{ii-1}.

\label{ii}

\begin{figure}
    \centering
    \includegraphics[width=1\linewidth]{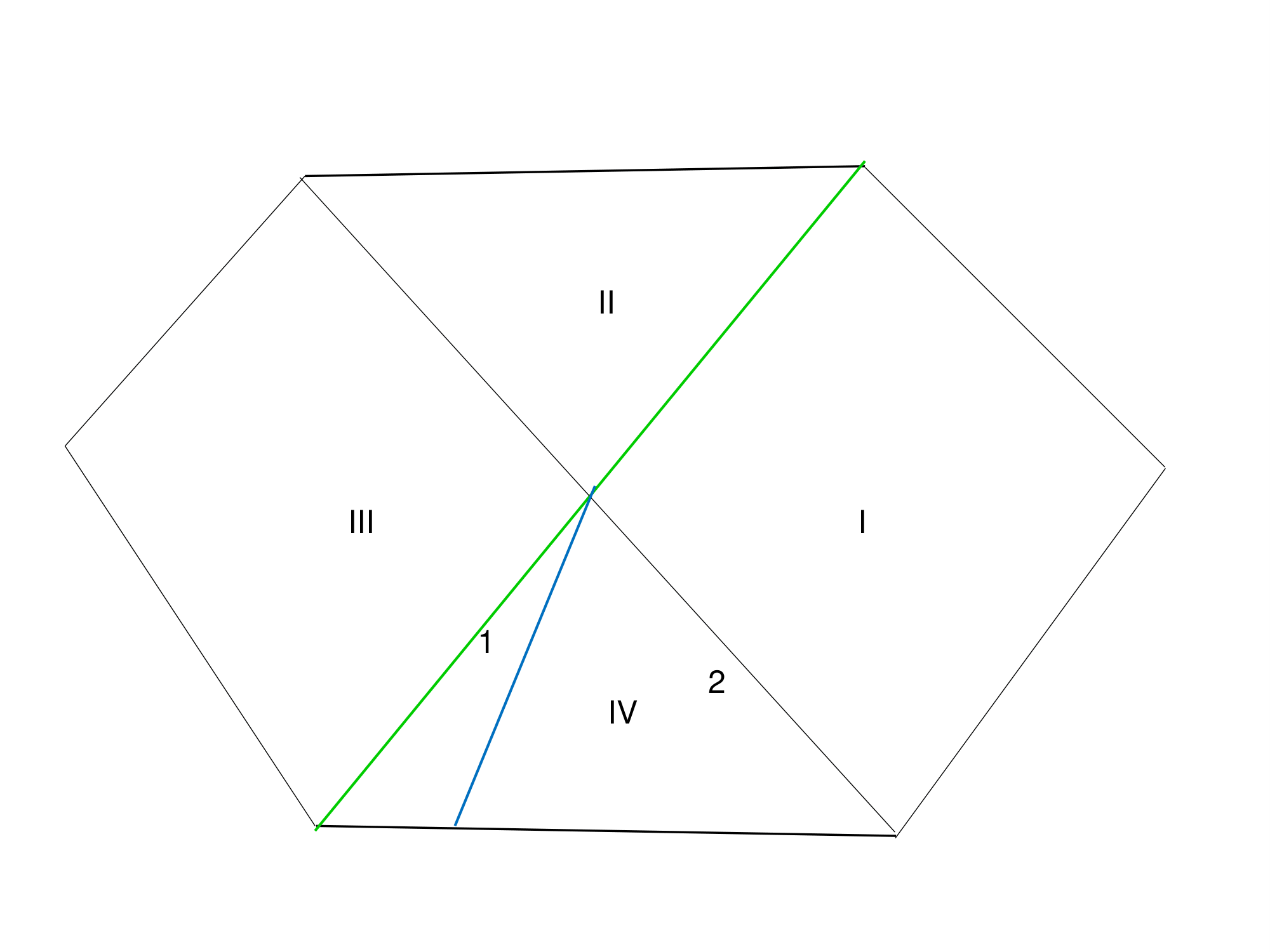}
    \caption{Scenario iii. Particle 1 is a photon moving along the horizon. Particle 2 is massive.}
    \label{20}
\end{figure}

(iii) We assume that from the very beginning both particles follow limiting
trajectories: a photon travels along the horizon, a massive particle moves
along the $RO$ trajectory. See Fig. (\ref{20}).

\label{iii}

There are also "mirror" scenarios in which propagation of particles changes
direction but there is no necessity to discuss them.

\section{Velocity vector and wave vector in time-like and space-like
Kruskal-Szekeres coordinates\label{kruskal}}

It is instructive to introduce, in addition to the light-like KS
coordinates, also the space-like $X$ and time-like $Y$ coordinates according
to 
\begin{equation}
U=Y-X,V=Y+X\text{.}  \label{ks}
\end{equation}%
Then, inside $WH$ (region $IV$)$,$ the KS non-angular coordinates $X,$ $Y$
are expressed via $T,$ $R$ coordinates ($T=r$, $R=t$) as follows:

\begin{equation}
Y=-(1-\frac{T}{2M})^{1/2}e^{T/4M}\cosh \frac{R}{4M}=-\sqrt{\frac{T}{2M}}%
\sqrt{-f}e^{T/4M}\cosh \frac{R}{4M},  \label{Y}
\end{equation}

\begin{equation}
X=-(1-\frac{T}{2M})^{1/2}e^{T/4M}\sinh \frac{R}{4M}=-\sqrt{\frac{T}{2M}}%
\sqrt{-f}e^{T/4M}\sinh \frac{R}{4M},  \label{X}
\end{equation}%
and the linear element is

\begin{equation}
ds^{2}=\frac{32M^{3}}{T}e^{-T/2M}\left( -dY^{2}+dX^{2}\right) +T^{2}d\Omega
^{2}.  \label{met}
\end{equation}%
One can check that%
\begin{equation}
Y^{2}-X^{2}=-\frac{Tf}{2M}e^{T/2M}\geq 0\text{,}  \label{yx}
\end{equation}%
as in this region $f=1-\frac{2M}{T}\leq 0$. On the P- horizon $V=0$, so $%
X=-Y $; on the F-horizon $U=0$ and $Y=X$.

Time-like geodesics (\ref{1}) are described by%
\begin{equation}
u=\sigma \frac{\varepsilon }{f}\partial _{R}+\sqrt{\varepsilon ^{2}-f}%
\partial _{T}\text{.}  \label{u}
\end{equation}%
If $\sigma =1$, they escape from region $IV$ crossing P-horizon and enter
region $I$. If $\sigma =-1$, they escape from region $IV$ \ crossing
F-horizon and enter region $III$. In coordinates $X$, $Y$ the four-velocity
reads

\begin{equation}
u=\frac{1}{4M}[(\sigma \frac{X\varepsilon }{f}+Y\frac{\sqrt{\varepsilon
^{2}-f}}{f})\partial _{Y}+(\sigma \frac{Y}{f}\varepsilon +X\frac{\sqrt{%
\varepsilon ^{2}-f}}{f})\partial _{X}].  \label{u+-}
\end{equation}

In particular, in the white hole region $IV$%
\begin{equation}
u=-\frac{1}{4M\left\vert f\right\vert }[(Y\sqrt{\varepsilon ^{2}+\left\vert
f\right\vert }+\sigma X\varepsilon )\partial _{Y}+(\sigma Y\varepsilon +X%
\sqrt{\varepsilon ^{2}+\left\vert f\right\vert })\partial _{X}].  \label{uxy}
\end{equation}

Using (\ref{yx}) and the fact that $Y<0$ here, one can check that in this
region $u^{Y}>0$, so the forward-in-time condition is satisfied.

Near the horizon 
\begin{equation}
u^{X}\approx \frac{\varepsilon }{4M\left\vert f\right\vert }(\sigma
\left\vert Y\right\vert -X)
\end{equation}%
Let $\sigma =1$ (so a particle can move from region $IV$ to region $I$). \
Then, $u^{X}>0$, a particle crosses the right horizon.

Let $\sigma =-1$. Then, as $\left\vert Y\right\vert >X$, $u^{X}<0$, a
particle crosses the left horizon.

In other words, the sign of $u^{X}=$sign of $\sigma $.

If we put $\varepsilon =0$ in (\ref{u+-}), we obtain the four-velocity of $%
RO $

\begin{equation}
u_{RO}=-\frac{1}{4M\sqrt{-f}}(Y\partial _{Y}+X\partial _{X}).  \label{uro}
\end{equation}%
For such an observer, eq. (\ref{uvt}) with $t=t_{0}=const$ is valid. If we
rewrite it in terms of variables $X$ and $Y$, we obtain that along the
trajectory 
\begin{equation}
Y=X\coth \kappa t_{0}\text{.}  \label{cot}
\end{equation}%
In the white hole region $Y<0$. Therefore, $t_{0}>0$ if $X<0$ and $t_{0}<$ 0
if $X>0.$ Then, using (\ref{yx}), (\ref{uro}) we have%
\begin{equation}
u_{RO}=\frac{1}{4M}\sqrt{\frac{T}{2M}}e^{T/4M}(\cosh \kappa t_{0}\text{, }%
\sinh \kappa t_{0}).  \label{ur0}
\end{equation}%
For radial light-like geodesics in region $IV$ we have two variants.

If a photon travels in the negative direction, so $dX<0$,

\begin{subequations}
\begin{equation}
k=\left( B\partial _{Y}-B\partial _{X}\right) .  \label{B}
\end{equation}%
Such a photon is trapped on the P- horizon.

If a photon travels in the positive direction, so $dX>0$,

\end{subequations}
\begin{equation}
k=\left( A\partial _{Y}+A\partial _{X}\right) .  \label{A}
\end{equation}%
Such a photon is trapped on the F- horizon.

As $Y$ is a time-like coordinate, $A>0$ and $B>0$ due to the forward-in-time
condition.

\section{Frequency registered near horizon and approach to bifurcation point 
\label{nearhor}}

Now, we are ready to find the frequency $\omega =-k_{\mu }u^{\mu }$ and
trace different limiting transitions.

\subsection{Scenario (i)}

At first, let us consider scenario (i) in which collision occurs on the
horizon with some small but nonzero $X_{c}>0$ and $Y\approx -X$. See Fig. %
\ref{Si}. As $\varepsilon >0$, in the vicinity of this horizon the terms of
order $1/f$ compensate each other, so $u$ remain finite as it should be
since the metric is regular in the KS coordinates. Taking into account (\ref%
{yx}) we have%
\begin{equation}
u_{+}\approx \frac{(X+Y)\varepsilon }{4Mf}(\partial _{Y}+\partial
_{X})\approx \frac{\varepsilon e}{8MX}(\partial _{Y}+\partial _{X}).
\end{equation}

Near the P-horizon $X>0$,

\textit{\ }%
\begin{equation}
\omega _{P}=\frac{2BM\varepsilon }{X_{c}}\text{.}
\end{equation}%
Thus approaching the BP $T\rightarrow 2M$, i.e. $X_{c}\rightarrow 0$, the
recorded frequency grows indefinitely so $\omega _{P}\rightarrow \infty $ in
agreement with \cite{along}. However, if the escaping geodesic crosses the
horizon at finite $X_{c}$ before reaching the BP,

\begin{equation}
\omega _{P}<\infty .
\end{equation}%
Similar result is obtained in an equivalent situation on the F-horizon, 
\begin{equation*}
\omega _{F}<\infty .
\end{equation*}

Hence, one also finds that these kinds of collisions are unbounded, energy
outcome can be indefinitely large, but could not be infinite.

\subsection{Scenarios (ii) and (iii)}

Let us consider scenario (ii). Without collision, the light-like geodesic
would cross the P-horizon and enter region $I$, so $\sigma =+1$. The
observer measuring frequency is $RO$. Then, It follows from (\ref{ur0}) that
the frequency registered by $RO$ equals%
\begin{equation}
\omega _{RO}=8A\frac{M^{3/2}}{\sqrt{2}T^{1/2}}e^{-T/4M}\exp (\sigma \kappa
t_{0})\text{,}  \label{ome}
\end{equation}%
where $\sigma =-1$ in the first case and $\sigma =+1$ in the second one. The
quantity $A$ depends on a collision point but has a well-defined horizon
limit. If a photon propagates along the horizon leg, it follows from the
geodesic equations that $A=const$ \cite{along}.

In the horizon limit $T\rightarrow 2M$%
\begin{equation}
\omega _{RO}=4\frac{A}{\sqrt{e}}\frac{M}{\sqrt{2}}\exp (\sigma \kappa t_{0})
\label{hor}
\end{equation}%
is finite. We can also start from the $RO$ and a photon propagating along
the horizon and obtain (\ref{hor}). In this sense, there is a smooth limit
from (ii) to (iii).

One can see that the scenarios (ii) and (iii) cannot be obtained from the
scenario (i) by the limiting transition $X_{c}\rightarrow 0$, they do it on
their own in a different way. That is because the Resting Observer, $RO$,
follows geodesics represented by $\varepsilon =0$, and those couldn't escape
the region $IV$, crossing the horizon $P$ or $F$, represented by $f=0$. In
other words, that is because the two limits, $\varepsilon =0$, $f=0$ do not
commute.

It is also worth noting that for $t_{0}>0$ and $\sigma =+1$, in the limit $%
t_{0}\rightarrow \infty $, $\omega _{RO}\rightarrow +\infty $ while for $%
\sigma =-1$ this limit gives us $\omega _{RO}\rightarrow 0$.

\section{Conclusions\label{concl}}

Thus we enumerated the set of possible scenarios of high energy particle
collisions in the complete space-time of the Schwarzschild black-white hole.
High-energy collision should occur near the horizon, where the metric
coefficient $f$ in (\ref{1met}) is small. This relies essentially on the
presence of a white hole and/or mirror regions in the full space-time
diagram. Also, it requires some kinematic conditions in the point of
collision. Possible set of scenarios of collisions with high energy outcome
contains the two main groups. The first one includes head-on-collisions,
i.e. those where geodesics do not pass through, do not originate at the
bifurcation point and have $\varepsilon =O(1)$. Then, $\gamma \sim
\left\vert f\right\vert ^{-1}$. The second group contains scenarios in which
one of the particles passes through or originates at the bifurcation point
i.e. this is a resting observer ($RO$), $\varepsilon =0$. Then, $\gamma \sim
\left\vert f\right\vert ^{-1/2}$. There is also an additional third group
when both particles have $\varepsilon =O(1)$ but collision occurs near the
bifurcation point. Such scenarios can be obtained by continuous
transformation from the first group and do not form a separate group on its
own.

Another issue concerns analysis of the principle of kinematic censorship for
collisions which becomes nontrivial in the vicinity of the bifurcation
point. We showed that, in spite of its seeming violation of it, this
principle remains valid even if one takes into account the discontinuous
nature of some limits. To this end, we discussed a rather tricky case of
collisions between a massive particle and a photon in the bifurcation point.
Three different limiting transitions have been considered. (i) Collision
occurs exactly on the horizon and the point of collision approaches the
bifurcation one. In this limit the trajectory of a massive particle
approaches that of $RO$. (ii) A\ massive particle is a $RO$ that meets a
photon in some intermediate point inside a white hole. Afterwards, this
point approaches the horizon. In doing so, the photon trajectory becomes
closer and closer to the horizon leg. (iii) A massive particle is a $RO$
trajectory and the photon moves along the leg of a horizon, so both
particles meet in the BP.

We saw that, symbolically, 
\begin{equation}
(iii)=\lim (ii),
\end{equation}%
with the frequency registered by such an observer remaining finite. However, 
$\lim (i)\neq (iii)$ and, moreover, 
\begin{equation}
\lim (i)=\infty
\end{equation}

Naive viewpoint could suggest that in the end of the procedure described in
(i) we could obtain an infinite blueshift (thus an infinite energy in the
center of mass frame). This would contradict the kinematic censorship
according to which in any event the energy cannot be infinite literally \cite%
{cens}. However, this does not occur since, as underlined above, the two
limiting transitions do not commute.

We hope that the developed approach will be useful for analysis and
classification of high-energy particle collisions near an outer and inner
horizons in more complex metrics like the Reissner-Nordstr\"{o}m one. In
particular, this will enable us to revisit and expand the results on
particle collisions inside the horizon known up to date.

\end{document}